\documentclass{iopjournal}

\usepackage{subfigure}
\usepackage{subcaption}
\usepackage{hyperref}

\begin{document}

\articletype{Paper} %	 e.g. Paper, Letter, Topical Review...

% \title{Validation of geometric metric-based predictions with experiments on TCV and multi-machine gyrokinetic simulations}
\title{Validation of geometric proxies using negative and positive triangularity plasmas with varying elongation on TCV and multi-machine gyrokinetic simulations}

% \author{A. Balestri$^{1,*}$\orcid{0009-0009-8309-7736}, J. Ball$^1$\orcid{0000-0002-4462-6350}, S. Coda$^{1}$\orcid{0000-0002-8010-4971}, O.Sauter${^1}$\orcid{0000-0002-0099-6675}, P. Mantica$^{2}$\orcid{0000-0001-5939-5244}, A. Mariani$^{2}$\orcid{0000-0003-0476-3825}, B. Labit$^1$\orcid{0000-0002-0751-8182}, R. J. J. Mackenbach$^1$\orcid{0000-0001-7056-0861}, B. Vanovac${^3}$\orcid{0000-0003-4031-9318} and the TCV team${^4}$}
\author{A. Balestri$^{1,2,*}$\orcid{0009-0009-8309-7736}, J. Ball$^1$\orcid{0000-0002-4462-6350}, S. Coda$^{1}$\orcid{0000-0002-8010-4971}, O. Sauter${^1}$\orcid{0000-0002-0099-6675}, P. Mantica$^{3}$\orcid{0000-0001-5939-5244}, A. Mariani$^{3}$\orcid{0000-0003-0476-3825}, B. Labit$^1$\orcid{0000-0002-0751-8182}, B. Vanovac${^4}$\orcid{0000-0003-4031-9318}, J. Hobirk$^{2}$\orcid{0000-0001-6605-0068} and the TCV team${^5}$}

\affil{$^1$École Polytechnique Fédérale de Lausanne (EPFL), Swiss Plasma Center (SPC), 1015 Lausanne, Switzerland}

\affil{$^2$Max-Planck-Institut für Plasmaphysik, Boltzmannstr. 2, 85748 Garching, Germany}

\affil{$^3$Istituto per la Scienza e la Tecnologia dei Plasmi, CNR, 20125 Milan, Italy}

\affil{$^4$Plasma Science and Fusion Center, Massachusetts Institute of Technology, Cambridge, MA 02139,
United States of America}

\affil{$^5$ See Theiler et al 2026 (https://doi.org/10.1088/1741-4326/ae6d12) for the TCV Team}

\affil{$^*$Author to whom any correspondence should be addressed.}

\email{alessandro.balestri@ipp.mpg.de}

\keywords{negative triangularity, gyrokinetic simulations, nuclear fusion, plasma physics, TCV}

\begin{abstract}
Negative Triangularity (NT) tokamak plasmas can achieve H-mode-like confinement while avoiding Edge Localized Modes, making them an attractive reactor scenario. Predicting in which configuration this benefit occurs requires understanding how magnetic geometry controls turbulent transport. Using a set of geometric proxies based on the poloidally averaged magnetic drift velocity, its poloidal variation, and Finite Larmor Radius (FLR) effects, this work validates their predictive power against gyrokinetic simulations and experiments. First, the proxies are used to predict how the relative benefit of NT over Positive Triangularity (PT) evolves with elongation $\kappa$, forecasting a detrimental effect of NT below $\kappa\sim0.75$--$0.8$. This prediction is confirmed by linear and nonlinear GENE simulations in both ITG- and TEM-dominated regimes. Additionally, a dedicated experimental elongation scan on TCV shows that NT plasmas at $\kappa^{\rm LCFS}=1.5$ have $\beta_N$ and $H_{98y2}$ roughly 1.5 times larger than PT counterparts, while no difference is observed at $\kappa^{\rm LCFS}=0.95$. The analysis is then extended to realistic ITG-dominated scenarios from DIII-D, AUG, and JET, including up-down asymmetric equilibria where conventional triangularity-based descriptions break down. In all three devices, simulations of NT equilibria exhibit substantially reduced nonlinear heat fluxes relative to PT (factors of 1.8, 1.6, and 2.1, respectively). A systematic swap of geometric coefficients between NT and PT equilibria shows that magnetic drifts and FLR effects are the dominant stabilizing mechanisms across all machines, with parallel dynamics playing a secondary role. These results establish the geometric proxies as reliable, computationally inexpensive predictors of NT transport benefits across diverse shapes and turbulence regimes.
\end{abstract}

\section{Introduction}

The present-day reference scenario for future tokamak fusion reactors relies on the High confinement (H-mode) regime. In this regime, when the power crossing the separatrix exceeds a certain threshold (i.e. the LH power threshold \cite{Martin_2008}), the plasma undergoes a sudden state transition and a transport barrier forms in a very narrow layer at the edge of the plasma. This transport barrier greatly reduces turbulence, enabling a strong steepening of the plasma pressure within the barrier itself. This structure is called the pedestal and lifts the whole plasma pressure radial profile, almost doubling the performance of the plasma. However, the promise of H-mode is tempered by two important factors. Firstly, the necessity of a steady minimum power flowing from the core through the separatrix to maintain H-mode creates the \textit{core-edge integration problem}. This minimum power inevitably flows to the plasma-facing components of the device, putting their integrity at risk. Thus, it is necessary to mitigate this power by means of impurity injection, which will radiate away some of the power in the form of electromagnetic radiation. However, if too much power is radiated away, H-mode is lost. The second major problem is the appearance of powerful instabilities called Edge Localized Modes (ELMs), which correspond to quasi-periodic relaxations of the pedestal and release large amounts of particles and energy into the plasma-facing components of the tokamak. In addition, the power released by an ELM is directly proportional to the amount of energy stored in the plasma, making ELMs extremely dangerous for larger devices. A single ELM could seriously damage the first wall of a power plant and compromise the steady production of energy \cite{Loarte_2014}. 

Alternative scenarios where ELMs can be avoided and the core-edge integration problem mitigated, while retaining the performance of an H-mode, are needed. Over the last decade, Negative Triangularity (NT) plasmas have emerged as a promising solution. They are characterized by a robust inhibition of H-mode \cite{Medvedev_2015,Merle_2017,Saarelma_2021,Nelson_2022,Sauter2023,Nelson_2024,Paz-Soldan_2021,Sauter2023_APS}, which prevents the formation of the pedestal and the subsequent ELMs, while achieving H-mode-like performance \cite{Austin_2019,Marinoni_2019,Fontana_2020,Coda_2022,Happel_2023,Thome_2024,Paz-Soldan_2024,Balestri_2024,Aucone_2024,Rienacker_2026} through a strong reduction of turbulent transport \cite{Marinoni_2009,Merlo_2015,merlo_jenko_2023,Balestri_2024,Balestri_2024_2,Merlo_ITG,Mariani_2024,Di_Giannatale_2024,Hoffmann_2025,Ulbl_2025,Balestri_2025,Becoulet_2026,DeGianni_2026}. This combination of features is attractive as a reactor scenario because it removes one of the most serious threats to the plasma-facing components of a reactor and makes the core-edge integration easier. Without the need to exceed the LH power threshold, it becomes possible to radiate away a large fraction of the power flowing to the walls by injecting impurities.

H-mode-like performance in NT plasmas can be achieved only if turbulent transport is sufficiently reduced. Thus, it is paramount to understand how the geometry of the flux surfaces affects turbulence. In recent years, different numerical works have shown that the beneficial effect of NT on transport depends on other geometrical parameters, e.g. magnetic shear \cite{merlo_jenko_2023,DeGianni_2026}, aspect ratio \cite{Balestri_2024_2} and shear of triangularity \cite{Merlo_ITG}.

It is clear that the parameter space that needs to be explored is vast. The main idea behind this work is to provide simple and reliable quantities that reflect the impact of the full magnetic geometry on turbulent transport, rather than relying on a restricted set of shaping parameters such as elongation and triangularity. In \cite{Balestri_2024_2}, we introduced a set of geometric proxies, based on the poloidally averaged magnetic drift velocity, its poloidal variation, and Finite Larmor Radius (FLR) effects, which capture the dominant mechanisms through which the magnetic geometry enters the gyrokinetic equations. %In appendix \ref{A} we recall the definition of these metrics, how they enter the gyrokinetic equation and how they are linked to turbulence reduction. 
These proxies have the advantage of being computable directly from any magnetic equilibrium, without requiring a decomposition into elongation, triangularity, or other parameters. This is particularly useful when the plasma shape is not up-down symmetric, as is often the case in present-day NT experiments \cite{Balestri_2024,Mariani_2024,Vanovac_2024,Vanovac_2026,Mantica_2026}. For these configurations the very definition of triangularity becomes ambiguous and comparison between different tokamaks is not trivial, as observed in \cite{Mantica_2026, Mariani_2026}. In contrast, our proxies remain well defined and retain a clear physical interpretation regardless of the up-down symmetry of the equilibrium.

%This is particularly useful for two reasons. First, as already shown in \cite{Balestri_2024_2} and further demonstrated in this work, negative triangularity alone is not a sufficient condition for reduced turbulent transport. Its beneficial effect depends strongly on other geometric and physical parameters. For instance, magnetic shear can change the impact of triangularity \cite{merlo_jenko_2023}, and in \cite{Balestri_2024_2} we showed that the aspect ratio dependence itself is regime-dependent: NT improves confinement relative to PT for ITG turbulence at any aspect ratio, whereas for TEM turbulence the benefit is only retained at large and conventional aspect ratio and is lost, and even reversed, at tight aspect ratio. With our metrics, it is no longer necessary to disentangle this kind of interplay between triangularity and other parameters case by case, since the impact of the overall shape can be assessed directly and qualitatively. Second, when the plasma shape is not up-down symmetric, as is increasingly relevant for present-day and future NT experiments, the very definition of triangularity becomes ambiguous. In contrast, our metrics remain well defined and retain a clear physical interpretation regardless of the up-down symmetry of the equilibrium.

The physical picture underlying these geometrical proxies was first developed in \cite{Balestri_2024_2}, where a simplified model of the toroidal Ion Temperature Gradient (ITG) instability was used to show that, in the large aspect ratio limit, NT stabilizes ITG turbulence through faster magnetic drifts with respect to a resonance condition set by the diamagnetic drift, combined with stronger FLR effects. That picture was shown to remain valid at conventional aspect ratio, but not at tight aspect ratio. Additionally, it did not extend cleanly to pure Trapped Electron Mode (TEM) turbulent regimes, where parallel dynamics and collisional effects were found to also play an important role. This left open the question of how well this framework applies to realistic scenarios and whether it can be used to make valid predictions. These are the two questions we address in this paper. We first use the proxies to predict how the benefit of NT relative to PT should vary with elongation, and we test this prediction with both gyrokinetic simulations performed with GENE \cite{GENE} and a dedicated experimental scan on the TCV tokamak. We then extend the validation to realistic ITG-dominated scenarios from DIII-D, AUG and JET, including equilibria that are not up-down symmetric, and we use GENE gyrokinetic simulations to disentangle the relative contributions of magnetic drifts, FLR effects and parallel dynamics to the observed differences between NT and PT.

The remainder of the paper is structured as follows. Section \ref{2} summarizes the findings of \cite{Balestri_2024_2} and introduces the proxies that will be used and tested in the rest of the paper. Section \ref{3} shows the results of a numerical investigation on the interplay between triangularity and elongation. Section \ref{4} shows the experimental results of an elongation scan with negative and positive triangularity in TCV, which confirmed the predictions of the previous section. Section \ref{5} shows a multi-machine gyrokinetic study of realistic scenarios dominated by ITG turbulence, spanning DIII-D, AUG and JET. Section \ref{6} draws the conclusions and provides a discussion.

\section{Theoretical model and geometrical proxies}\label{2}
In this section we provide a short introduction to the GENE code, which has been used for all the simulations presented in this work. We also introduce the definitions of the three proxies that will be used in this work, how they enter the gyrokinetic equation and how they are linked to turbulence.

GENE is a physically comprehensive Eulerian gyrokinetic code that solves the Vlasov-Maxwell equations discretized on a 5-dimensional (5D) grid. The code can be run in a linear mode, if the nonlinear term of the GK equation is neglected, or in a nonlinear mode if the term is retained. The former is much less computationally expensive and is especially useful to develop an understanding of the nature of the turbulence that dominates in the system. The latter is much more expensive but is needed to have a physically accurate model of micro-turbulence. Indeed, one must consider the nonlinear interaction for turbulence to saturate and to assess the transport level. The real space is parametrized by a 3D set of non-orthogonal field-aligned coordinates
\begin{equation}
    \begin{cases}
        x=x(\psi)\\
        y=C_y\left(q(\psi)\theta-\varphi\right),\\
        z=\theta
    \end{cases}
    \label{coordinates}
\end{equation}
which correspond respectively to the radial, binormal, and parallel (to B) coordinates. Here $\psi$ is the radial coordinate, $\varphi$ the toroidal angle, $\theta$ the poloidal angle, $q$ the safety factor and $C_y$ is a constant. The remaining 2 coordinates of the 5D discretized grid, $v_\parallel$ and $\mu$, parametrize the velocity space and correspond to the parallel velocity and the magnetic moment, respectively. In the flux tube domain used in this paper, the kinetic profiles are Taylor expanded around the flux surface of interest so that their values and gradients are constant. This choice is justified if the radial turbulence scale is much smaller than the radial variation of equilibrium quantities (machine scale). This is the case for most medium-size tokamaks, and is especially true for reactor-scale devices. Therefore, the real-space simulation domain corresponds to a small rectangle extended in $x$ and $y$ at the outboard midplane, which follows the magnetic field lines along $z$. Periodic boundary conditions are used for the perpendicular coordinates $x$ and $y$, while pseudo-periodic boundary conditions that account for magnetic shear are applied in $z$. GENE has various options to reconstruct the magnetic equilibrium at a specific flux surface. For all the simulations presented in this work, we used the local equilibrium Miller model \cite{miller,Candy_2009} for up-down symmetric equilibria and the Fourier version \cite{Candy_2009} for up-down asymmetric equilibria.

\subsection{Geometry dependence in the gyrokinetic equations}
The gyrokinetic Vlasov equation for a species $\sigma$ solved by GENE in the local limit can be written as
\begin{equation}
    \begin{split}
        \frac{\partial g_{1\sigma}}{\partial t}&=\frac{B_0}{\mathcal{C}B_{0\parallel}^*}\left(\frac{1}{n_{0\sigma}} \partial_x n_{0\sigma}+\left( \frac{m_\sigma v_{\parallel\sigma}^2}{2T_{0\sigma}}+\frac{\mu_\sigma B_0}{T_{0\sigma}}-\frac{3}{2}\right)\frac{1}{T_{0\sigma}}\partial_x T_{0\sigma}\right)F_{0\sigma}\partial_y\bar{\xi}_1\\&
        -\frac{B_0}{\mathcal{C}B_{0\parallel}^*}\left( \partial_x\bar{\xi}_1\Gamma_{y\sigma}-\partial_y\bar{\xi}_1\Gamma_{x\sigma}\right)\\&
        +\frac{B_0}{\mathcal{C}B_{0\parallel}^*}\left( \frac{m_\sigma v_{\sigma\parallel}^2+\mu_\sigma B_0}{q_\sigma B_0} \right)(v_{Dx}\Gamma_{x\sigma} + v_{Dy}\Gamma_{y\sigma})\\&
        -\frac{B_0}{\mathcal{C}B_{0\parallel}^*}\frac{4\pi v_{\parallel\sigma}^2}{q_\sigma B_0^2}\partial_x P_{0\sigma}\Gamma_{y\sigma}\\&
        -\frac{\mathcal{C}}{JB_0}v_{\parallel\sigma}\Gamma_{z\sigma}\\&
        +\frac{\mu_\sigma}{m_\sigma}\frac{\mathcal{C}}{JB_0}\partial_z B_0\frac{\partial \bar{f}_{1\sigma}}{\partial v_\parallel}\\&
        +\langle C(\bar{f}_{1\sigma})\rangle,
    \end{split}
    \label{gyro_final}
\end{equation}
where $n_{0\sigma}$ is the equilibrium density, $T_{0\sigma}$ the equilibrium temperature, $P_{0\sigma}$ the equilibrium pressure, $m_\sigma$ the mass of the particle $\sigma$, $\mu_\sigma$ is the magnetic moment, $q_\sigma$ the particle charge, $v_{\parallel\sigma}$ the parallel velocity, $B_0$ the equilibrium magnetic field, $\textbf{B}_0^*=\textbf{B}_0+\frac{B_0}{\Omega}v_\parallel\boldsymbol{\nabla}\times\textbf{b}+\boldsymbol{\nabla}\times(\textbf{b}\bar{A}_{1\parallel})$ is the effective magnetic field, $B_{0\parallel}^*=\textbf{B}_0^*\cdot\textbf{b}$ is the parallel component of the effective magnetic field, $\textbf{b}=\textbf{B}_0/B_0$ is the unit vector parallel to the magnetic field, $\bar{\xi}_1=\bar{\phi}_1-v_\parallel\bar{A}_{1\parallel}+\frac{\mu}{q}\bar{B}_{1\parallel}$ is the gyroaveraged generalized potential, $\phi_1$ the perturbed electrostatic potential, $A_{1\parallel}$ the perturbed vector potential, $B_{1\parallel}$ the parallel component of the perturbed magnetic field, $g_{1\sigma}=\bar{f}_{1\sigma}-q_\sigma v_{\parallel\sigma}\bar{A}_{1\parallel}\frac{F_{0\sigma}}{T_{0\sigma}}$ is the modified perturbed particle distribution function, $f_{1\sigma}$ is the perturbed distribution function and
\begin{equation}
\begin{split}
    \mathbf{\Gamma}_\sigma=&\boldsymbol{\nabla}g_{1\sigma}+\frac{q_\sigma}{T_{0\sigma}}F_{0\sigma}\boldsymbol{\nabla}\bar{\xi}_1\\+&q_\sigma \bar{A}_{1\parallel}\frac{v_{\parallel\sigma}}{T_{0\sigma}}F_{0\sigma}\left(\frac{1}{n_{0\sigma}}\boldsymbol{\nabla}n_{0\sigma}+\left( \frac{m_\sigma v_{\parallel\sigma}^2}{2T_{0\sigma}}+\frac{\mu_\sigma B_0}{T_{0\sigma}}-\frac{3}{2}\right)\frac{1}{T_{0\sigma}}\boldsymbol{\nabla}T_{0\sigma}-\frac{\mu_\sigma}{T_{0\sigma}}\boldsymbol{\nabla}B_0   \right).
    \end{split}
\end{equation}

Equation \eqref{gyro_final} is coupled to the gyrokinetic Maxwell equations \cite{3} through the evolution of the electrostatic potential and the parallel component of the vector potential. In equation \eqref{gyro_final} we can identify all the terms on the right hand side that contribute to the evolution of the gyrokinetic distribution function $g_{1\sigma}$ and identify those that depend on the geometry of the flux surface. It must be noted that in the flux tube model, the term $B_0/(\mathcal{C}B_{0\parallel}^*)$ is a constant. The first term is the drive term. It does not depend on the magnetic geometry of the flux surface, except for the $J_0(k_\perp\rho)$ term that comes from the gyroaverage operation applied to the generalized potential $\xi_1$, which will be explained in the following. The second term is the nonlinear term, which determines the nonlinear evolution of the system and depends on geometry through the $J_0(k_\perp\rho)$ term. In GENE, it is possible to solve the gyrokinetic equation with or without this term depending on whether the user wants to study only the linear physics or include also the nonlinear dynamics. The third term is the magnetic drift term, which strongly depends on the magnetic geometry through $v_{Dx}$ and $v_{Dy}$, i.e. the radial and binormal components of the magnetic drift velocity. These are some of the metrics that strongly enter our model and are defined as
\begin{equation}
    v_{Dx}=-\left(\partial_y B_0 + \frac{\gamma_2}{\gamma_1} \partial_z B_0\right),
\end{equation}
\begin{equation}
    v_{Dy}=\left(\partial_x B_0 - \frac{\gamma_3}{\gamma_1} \partial_z B_0\right),
\end{equation}
where
\begin{equation}
    \gamma_1 = g^{xx}g^{yy} - g^{yx}g^{xy} \qquad
\gamma_2 = g^{xx}g^{yz} - g^{yx}g^{xz} \qquad
\gamma_3 = g^{xy}g^{yz} - g^{yy}g^{xz}
\end{equation}
and the terms $g^{ij}=\boldsymbol{\nabla}x^i\cdot\boldsymbol{\nabla}x^j$ are the terms of the metric tensor $g^{ij}$. The fourth term in equation \eqref{gyro_final} is the pressure term, which depends on geometry through the gyroaverage in $\Gamma_y$. The fifth term is the parallel streaming term, which depends on magnetic geometry through $1/(JB_0)$ and $\partial_zB_0$ in $\Gamma_z$. The sixth term is the mirror, which depends on geometry through $\partial_zB_0/(JB_0)$. Finally, the last term represents collisions. The linearized Landau-Boltzmann operator implemented in GENE and used in this work acquires an explicit geometry dependence because velocity space is parametrized by $(v_\parallel,\mu)$ rather than by $(v_\parallel,v_\perp)$. Since the magnetic moment $\mu=m v_\perp^2/(2B_0)$ is defined through the local field strength, evaluating the relative speed entering the collision kernels, $v=\sqrt{v_\parallel^2+\mu B_0}$, requires $B_0$ at each point along the field line. As a consequence, the collisional friction and diffusion coefficients, together with the momentum- and energy-conserving corrections of the field-particle operator, vary with $B_0$ in the same way as the mirror term above, effectively coupling collisions to the local magnetic field strength. In \cite{Balestri_2024_2}, we observed that this geometric dependence of the collisional operator does not play a major role in ITG turbulence, but can become important in TEM-dominated turbulent regimes. Geometry enters the gyrokinetic equations also through the gyroaverage operations. In the GK equation, we have
\begin{equation}
    \bar{G}_{1}=\frac{1}{2\pi}\int d\alpha G_{1}(\textbf{X}+\pmb{\rho}),
    \label{gyroaverage1}
\end{equation}
while in the field equations (GK Maxwell's equations) we have gyroaverages of the form
\begin{equation}
    \langle G_1(\textbf{x}-\pmb{\rho})\rangle=\frac{1}{2\pi}\int d\alpha G_{1}(\textbf{x}-\pmb{\rho}),
    \label{gyroaverage2}
\end{equation}
where $G_1$ is a generic fluctuating quantity appearing in the GK equations. In the local limit we apply periodic boundary conditions, which allow us to use a Fourier representation for all the fluctuating quantities in both perpendicular directions. It can be shown that the two gyroaverages in equations \eqref{gyroaverage1} and \eqref{gyroaverage2} are equivalent and can be expressed as
\begin{equation}
    \bar{G}_1=\langle G_1\rangle=J_0(k_\perp\rho)G_1,
\end{equation}
where $J_0$ is the zeroth order Bessel function and
\begin{equation}
    k_\perp=\sqrt{g^{xx}k_x^2+2g^{xy}k_xk_y+g^{yy}k_y^2},
\end{equation}
is the perpendicular wavenumber. The mechanism described by $J_0(k_\perp\rho)$ is called the Finite Larmor Radius (FLR) effect and depends on geometry through the metric tensor elements $g^{xx}$, $g^{xy}$ and $g^{yy}$. Stronger FLR effects imply a lower value of $J_0$, which, as it appears in the GK equations as a simple multiplication factor in front of the perturbed quantities, tends to stabilize turbulence. FLR effects represent the fact that GK theory models particles with a non-zero gyroradius, which, depending on the size of the turbulent eddies, will respond in different ways.

\subsection{Physical picture and definition of geometrical proxies}

Inspecting equation \eqref{gyro_final} and the GK Maxwell's equation, we observe that the geometry of the flux surface enters in the GK local description essentially through the magnetic drift velocities $v_{Dx}$ and $v_{Dy}$, the zeroth order Bessel function $J_0(k_\perp\rho)$ that models the FLR effects and the terms $1/(JB_0)$ and $\partial_zB_0/(JB_0)$, which influence respectively the parallel streaming and mirror terms. In \cite{Balestri_2024_2}, in simplified regimes, i.e. dominated by pure-ITG and pure-TEM turbulence, at large and conventional aspect ratio we showed that magnetic drift and FLR effects play the most significant role in determining the impact of shaping on turbulence. Specifically, we observed that the stabilization of turbulence due to NT geometry is mostly due to stronger FLR effects and faster magnetic drifts that vary significantly along the poloidal angle around the outboard midplane. We explained why a fast, rapidly varying magnetic drift velocity stabilizes ITG and TEM turbulence by observing that this mechanism makes the ions and trapped electrons drift faster than the underlying wave, whose phase velocity is set by the diamagnetic drift velocity. This prevents them from coupling properly and thus reducing the drive of the instability. This interpretation is consistent with previous works \cite{Beer,Marinoni_2009,merlo_jenko_2023}. The stabilizing effect of stronger FLR effects in NT can be understood by considering that the NT geometry compresses the flux surfaces around the outboard midplane. This is reflected in a larger $g^{yy}$ in a region around the outboard midplane, which then translates into modes with a shorter perpendicular wavelength than in PT for the same pair ($k_x$, $k_y$). Therefore, since the gyroradius does not change significantly between NT and PT, a particle will sample more independent field variations within one gyroorbit, which thus increases the effect of the gyroaverage. This mechanism is also consistent with previous work \cite{Marinoni_2009,merlo_jenko_2023}.

In order to define simple geometrical proxies that reflect these physical properties, these effects can be modeled by taking the flux surface average $\langle...\rangle$ of the quantities $v_{Dy}$, $dv_{Dy}/dz$ and $g^{yy}$. The choice of focusing on $v_{Dy}$ and $g^{yy}$ has been justified in \cite{Balestri_2024_2}. Here we recall that in typical conditions, turbulence peaks for modes with a radial wavenumber $k_x\rho_i=0$. Therefore, all the geometrical terms multiplied by $k_x$ can be neglected at first order and only $v_{Dy}$ and $g^{yy}$ survive and play a significant role. The flux surface average of a generic function $A$ of the straight-field line poloidal angle is defined as
\begin{equation}
    \langle A\rangle=\frac{\int_{-\theta_B}^{\theta_B}dz\;A\, J}{\int_{-\theta_B}^{\theta_B}dzJ},
\end{equation}
where $z$ is the straight-field line angle used in GENE, $J$ the Jacobian and $\theta_B$ is the angle used to define the boundaries of the average. Because of the ballooning nature of ITG and TEM modes, the most important region for the stability of these modes is the outboard midplane. Therefore, to keep the computations of the geometrical proxies simple and retain the ballooning nature of these instabilities, we use in the rest of the paper $\theta_B=[-1,1]$, with $\theta_B$ expressed in radians. This choice is arbitrary, but well justified by the fact that typical ITG and TEM modes do not extend significantly beyond this angle. In addition, we observed that considering a larger or smaller interval for $\theta_B$ does not significantly change the trends in the geometrical proxies. This is due to the fact that the Jacobian peaks at the outboard midplane and decays sharply as one moves toward the inboard midplane, thus giving more relevance to the poloidal angles closest to the outboard midplane.

Now, as we observed that a strong stabilization of ITG and TEM modes comes from strong FLR effects and fast magnetic drifts that change strongly along the outboard midplane, we expect the stabilization of turbulence to be correlated with large values of $\langle |dv_{Dy}/dz|\rangle$ and $\langle g^{yy}\rangle$, and by strongly negative values of $\langle v_{Dy}\rangle$. In the following, we test this picture with realistic gyrokinetic simulations and experiments on the TCV tokamak.

We conclude by mentioning that a similar model, based on the geometric dependence of the magnetic drift velocities and FLR effects, is also being developed for stellarators \cite{Roberg-Clark_Xanthopoulos_Plunk_Stroteich_2026}.

\section{Numerical investigation of elongation and triangularity in ITG and TEM regimes}\label{3}
% \section{Numerical and experimental results for an elongation scan with negative and positive triangularity in TCV}\label{2}

\begin{figure}[htbp]
    \centering
    \includegraphics[width=\textwidth]{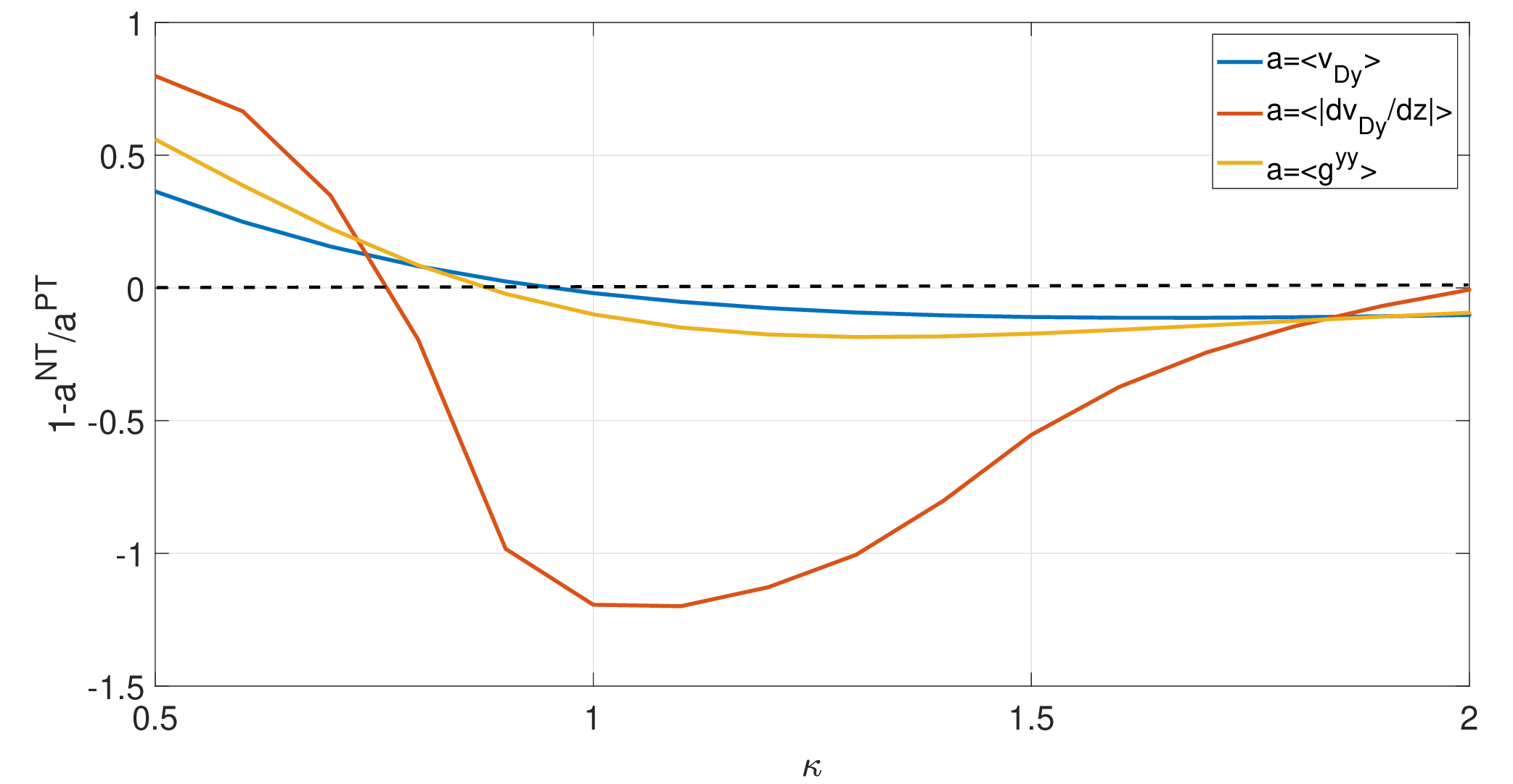}
    \caption{Relative differences of the poloidally averaged magnetic drift $\langle v_{Dy}\rangle$, absolute value of the poloidal variation of the magnetic drift $\langle |dv_{Dy}/dz|\rangle$ and FLR effects $\langle g^{yy}\rangle$ in NT and PT geometry as functions of elongation. The poloidal averages have been performed around the outboard midplane, i.e. $\theta=[-1,1]$.}
    \label{metrics_kappa}
\end{figure}

To validate the picture described in the previous section, we used the geometrical proxies to find realistic scenarios where NT is expected to be worse than PT. We considered the equilibrium of the TCV discharge \#68145 (discussed in \cite{Balestri_2024_2}) and we reconstructed the $\rho_{tor}=0.8$ surface with the Miller parametrization. We then flipped the sign of triangularity and its shear, and we artificially changed the elongation and its shear (following the formula $s_\kappa=(\kappa-1)/\kappa$). Finally, we computed the proxies for the NT and PT scenarios at different elongations. Figure \ref{metrics_kappa} shows the relative difference between NT and PT of the magnetic drift velocity, the poloidal variation of the magnetic drift and the FLR effects, poloidally averaged around the outboard midplane.

According to section \ref{2}, one would expect NT to perform better than PT in the range $\kappa=[1.0-2.0]$, i.e. where the relative differences of \textit{all} the quantities are negative. On the other hand, we would expect NT to have worse transport than PT in the range $\kappa=[0.5-0.8]$, where the relative difference of all three proxies is positive. It is interesting to notice that the absolute relative difference in $\langle |dv_{Dy}/dz|\rangle$ between NT and PT increases going from $\kappa=2.0$ to $\kappa=1.0$, while it decreases very rapidly and inverts for smaller values of elongation. On the contrary, the difference in $\langle v_D\rangle$ and FLR effect decreases monotonically going towards low values of elongation. Therefore, while it is clear that we expect NT to be worse than PT for $\kappa<0.8$ and better than PT for $\kappa>1$, it is hard to predict if the difference between NT and PT will increase or decrease in the remaining range of elongation. This will depend on the importance of the variation of the magnetic drift compared to $v_D$ and FLR effects. As a side note, we remark that a similar change in the beneficial effect of NT at different elongation for other magnetic equilibria has already been observed in \cite{Balestri_2024_2,Mackenbach_2023}.

To test this prediction, which is entirely based on the trend observed in the geometric coefficients, we performed gyrokinetic simulations and experiments on TCV.

We start by performing linear and nonlinear GK simulations of TEM and ITG scenarios. The TEM scenario is the realistic TEM-dominated TCV \#68145 scenario already described in \cite{Balestri_2024_2}. The ITG scenario is still based on the TCV \#68145 experiment, but we slightly modified the gradients to make ITG dominant. We call both of the cases realistic because the simulations do not have only one type of instability and they retain the effect of collisions, electromagnetic effects and the electrons are treated kinetically. Table \ref{elongation_input} reports the reference parameters used in the simulations.

\begin{table}[htbp]
    \centering
    \begin{tabular}{l|cc}
    % \hline
        &TCV \#68145 TEM & TCV \#68145 ITG  \\ \hline
        $R/L_{Te}$&14.54 & 8.48  \\ 
        $R/L_{Ti}$&8.48 & 14.54 \\ 
        $R/L_{ne}$&6.01 & 6.01  \\
        $T_e$ [keV]&0.41 & 0.41 \\ 
        $T_i/T_e$&0.45 & 1.0  \\ 
        $n_e\,[10^{19}m^{-3}]$& 1.95 & 1.95 \\ 
        $\beta [\%]$ &0.16 & 0.16 \\
        \hline
        $A$& 5 & 5 \\ 
        $q$&2.01 & 2.01 \\ 
        $\hat{s}$&1.6 & 1.6 \\
        $\kappa$ & 1.37 & 1.37 \\
        $|\delta|$ &0.16 & 0.16 \\
        $\zeta$ & -0.07 & -0.07 \\
        $s_\kappa$ & 0.19 & 0.19\\
        $|s_\delta|$& 0.33 & 0.33 \\
        $s_\zeta$ &-0.21 & -0.21 \\
        $\partial_rR$ &-0.23 & -0.23 \\\hline
    \end{tabular}
    \caption{\label{elongation_input} Logarithmic gradients of electron temperature $R/L_{Te}$, ion temperature $R/L_{Ti}$, electron density $R/L_{ne}$, as well as the ion-electron $T_i/T_e$ temperature ratio, electron density $n_e$, the local electron $\beta$ and geometric parameters needed to specify the Miller equilibrium.}
\end{table}

To perform our simulations we used the Miller parametrization to model the target flux surface and change the elongation artificially, producing an equilibrium that is still a local solution of the Grad-Shafranov equation. The sign of triangularity and its shear have been flipped to create a PT counterpart while the gradients have been kept fixed. The numerical convergence of all simulations has been tested by resolution studies. For linear simulations we used the resolution $(n_x,n_z,n_v,n_\mu)=(48,48,40,12)$. For nonlinear simulations we used $L_x\sim180\rho_i$ and $L_y\sim250\rho_i$ for the radial and binormal widths of the simulation box with $256$ $k_x$ and $48$ $k_y$ modes. For the parallel direction $z$ we used $48$ grid points. The size and resolution of velocity space have been kept fixed: $L_v=3v_{th,\sigma}$, $L_\mu=9\mu B_0/T_\sigma$, $n_v=42$ and $n_\mu=12$, where $v_{th,\sigma}=\sqrt{2T_\sigma/m_\sigma}$ is the thermal velocity, $B_0$ the toroidal magnetic field at the center of the flux surface of interest, $m_\sigma$ the reference mass and $T_\sigma$ the reference temperature.

\subsection{Linear simulations}
\begin{figure}[htbp]
    \centering
    \includegraphics[width=\textwidth]{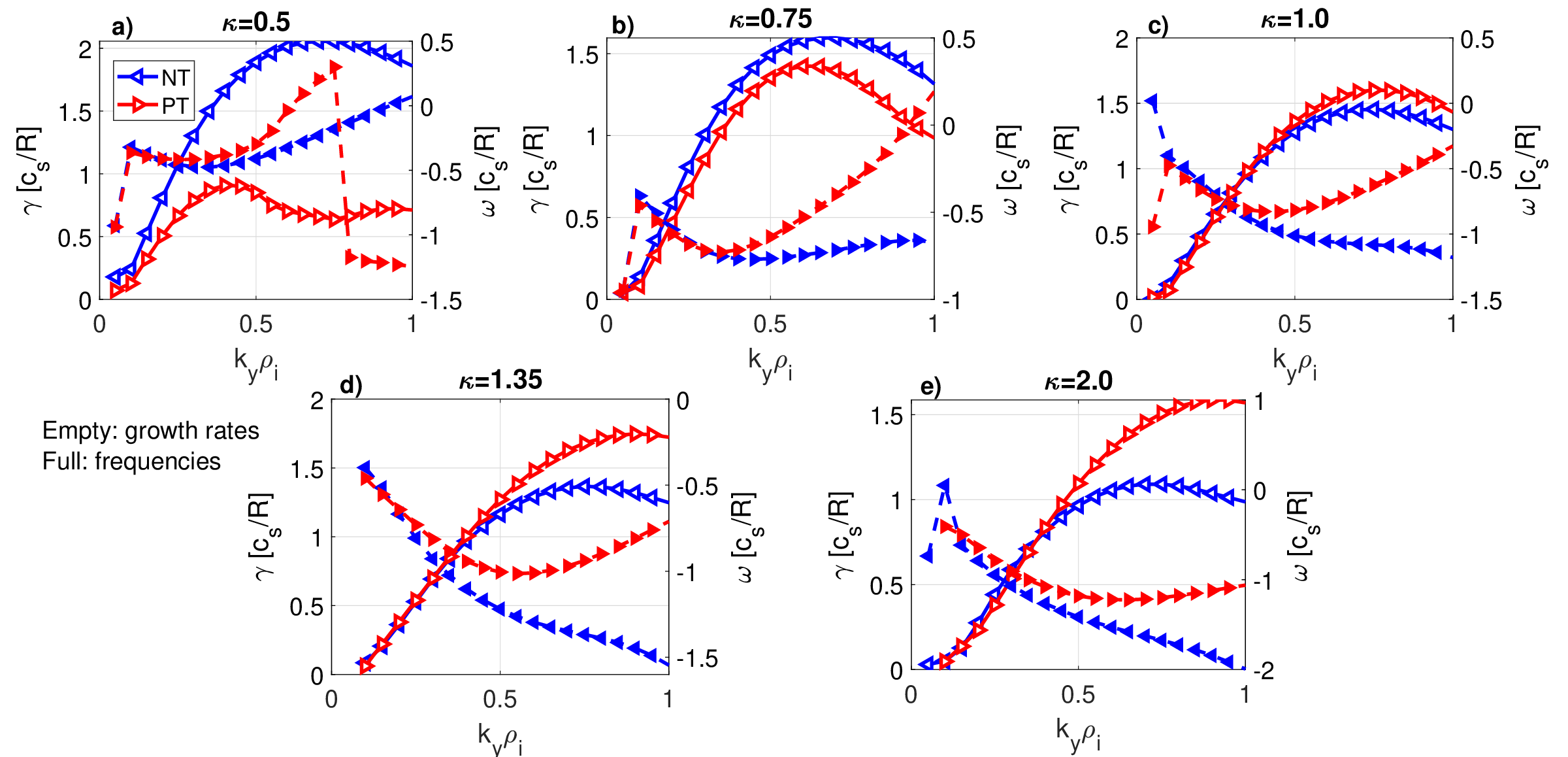}
    \caption{Growth rate $\gamma$ (empty symbols) and frequency $\omega$ (full symbols) of the most unstable modes as functions of the binormal wavenumber $k_y\rho_i$ at different elongations for NT (blue) and PT (red) TEM-dominated cases.}
    \label{gamma_omega_TEM_scankappa}
\end{figure}

\begin{figure}[htbp]
    \centering
    \includegraphics[width=\textwidth]{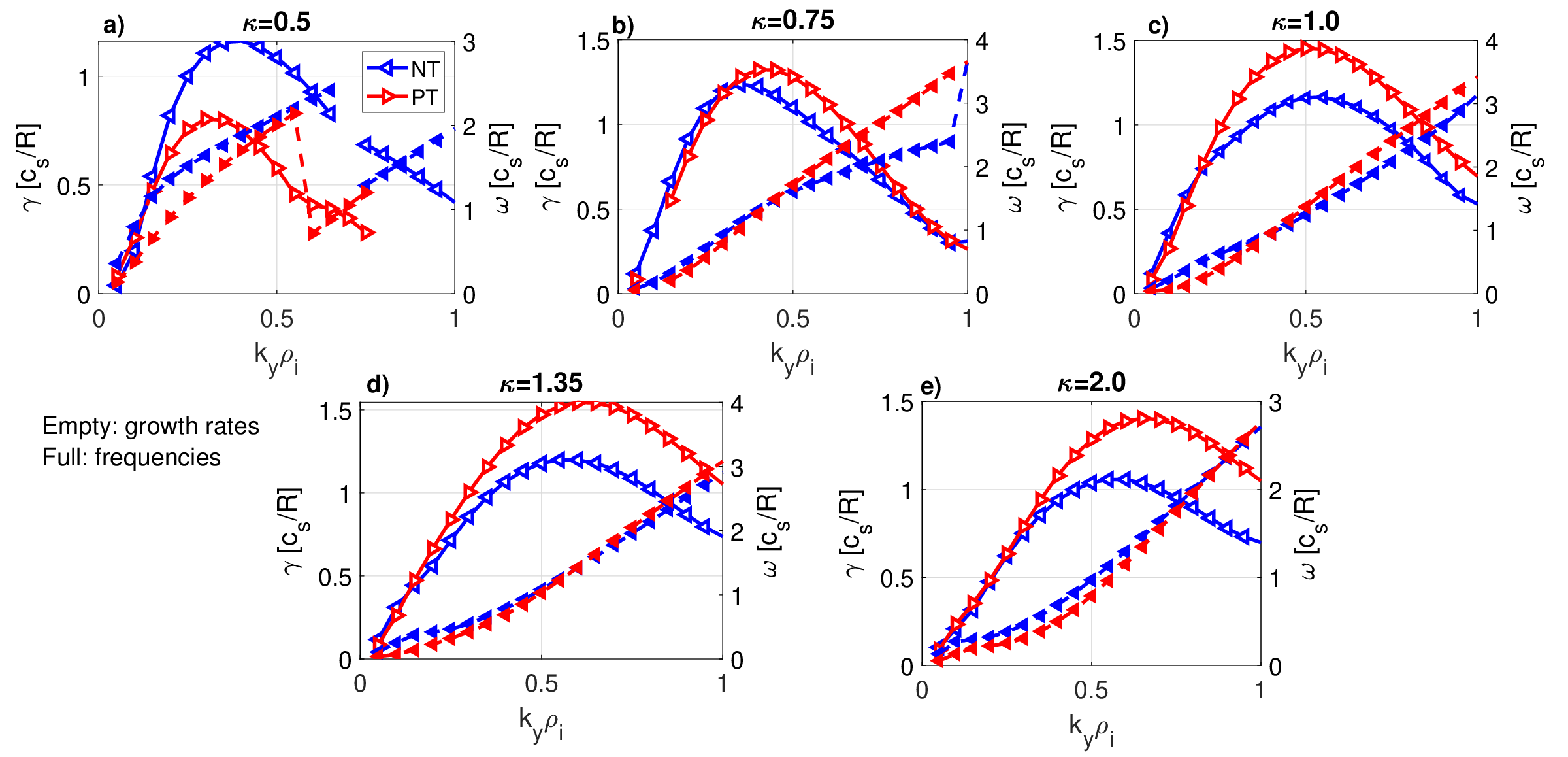}
    \caption{Growth rate $\gamma$ (empty symbols) and frequency $\omega$ (full symbols) of the most unstable modes as functions of the binormal wavenumber $k_y\rho_i$ at different elongations for NT (blue) and PT (red) ITG-dominated cases.}
    \label{gamma_omega_ITG_scankappa}
\end{figure}

To address the impact of elongation we started by performing linear simulations of the TEM and ITG scenario at the ion scale. Figure \ref{gamma_omega_TEM_scankappa} shows the results for the TEM-dominated scenarios at different elongations and figure \ref{gamma_omega_ITG_scankappa} shows the analogous results for the ITG case. In each plot, for a given value of $\kappa$ we display the growth rates and frequencies as functions of the binormal wave number $k_y\rho_i$ for the NT and PT cases. The frequency confirms that we have the expected turbulent regime, i.e. TEM for the TEM-dominated case and ITG for the ITG-dominated one. Looking at the growth rates, we observe the same trend in TEM and ITG. Specifically for $\kappa=0.5$ and $\kappa=0.75$, NT has stronger modes than PT, with the difference being bigger for lower elongation. On the contrary, when $\kappa>1$, NT is more stable than PT. At least linearly, this confirms our predictions based on the trend of the geometric coefficients.

\begin{figure}[htbp]
    \centering
    \includegraphics[width=\textwidth]{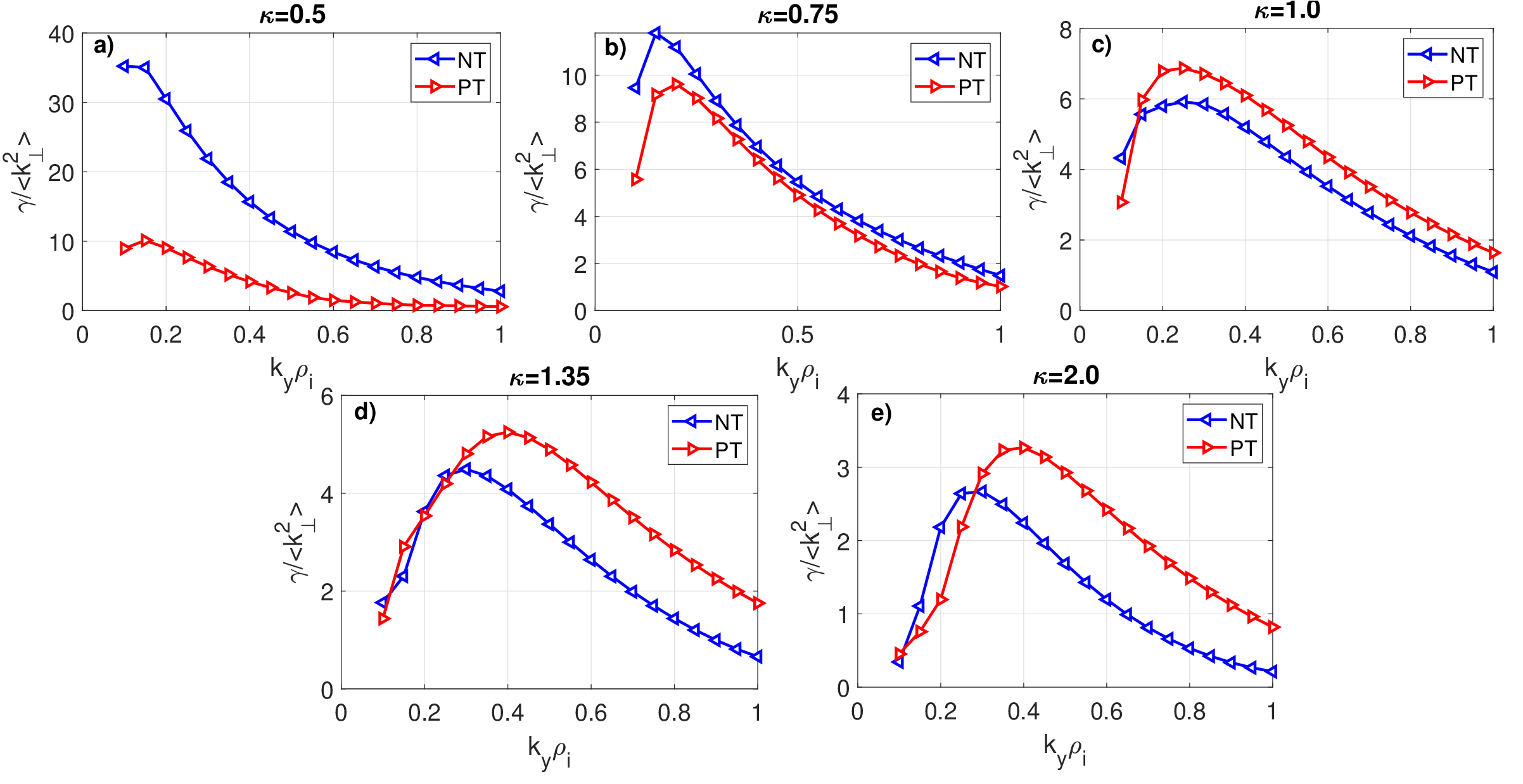}
    \caption{QL weights $\gamma/{\langle k_\perp^2\rangle}$ as a function of the binormal wavenumber $k_y\rho_i$ at different elongations for NT (blue) and PT (red) TEM-dominated cases.}
    \label{QL_TEM_scankappa}
\end{figure}

\begin{figure}[htbp]
    \centering
    \includegraphics[width=\textwidth]{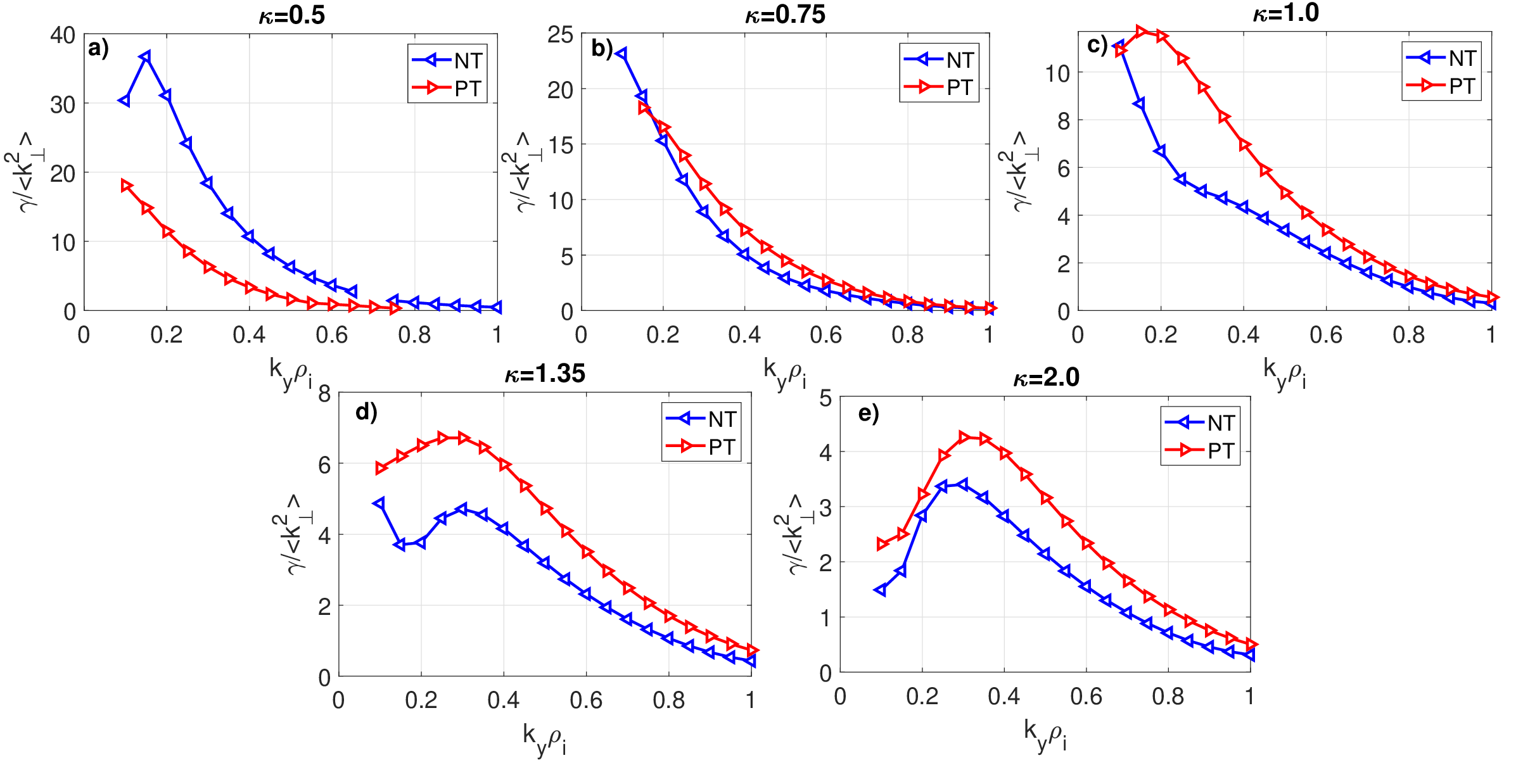}
    \caption{QL weights $\gamma/{\langle k_\perp^2\rangle}$ as a function of the binormal wavenumber $k_y\rho_i$ at different elongations for NT (blue) and PT (red) ITG-dominated cases.}
    \label{QL_ITG_scankappa}
\end{figure}

To have a preliminary understanding of how the different linear stability of the different scenarios will translate into different contributions to the nonlinear spectra, we used a simple mixing length approach to estimate the saturation level of turbulent transport. This approach is well described in \cite{Jenko_2005,Casati_2009,Mariani_QL}, where the QuasiLinear (QL) heat flux is written as
\begin{equation}
    Q^{QL}=A_0\sum_{k_y}\frac{\gamma(k_y)}{\langle k_\perp^2\rangle}Q_{ky}^L,
\end{equation}
where $\gamma(k_y)$ is the linear growth rate of each mode, $<k_\perp^2>$ is the flux surface averaged squared perpendicular wavenumber and $Q_{ky}^L$ is the linear heat flux normalized to the absolute value of the electrostatic potential squared at the outboard midplane. For more details, the reader is referred to \cite{Jenko_2005,Casati_2009,Mariani_QL}. In principle, $A_0$ is a shape dependent quantity. For this reason, we focus our attention only on how the QL weights $\frac{\gamma(k_y)}{\langle k_\perp^2\rangle}$ change with respect to a change in geometry. The study of this quantity is relevant because the diffusion coefficients scale as the QL weights \cite{Jenko_2005}. In figures \ref{QL_TEM_scankappa} and \ref{QL_ITG_scankappa} we show the QL weights for the TEM and ITG cases respectively as functions of the binormal wavenumber. The trends in the QL weights are consistent with the linear results. We observe that, for both TEM and ITG cases NT has larger QL weights across the full ion scale for $\kappa=0.5$. The QL weights are larger in NT $\kappa=0.75$ for $k_y\rho_i<0.15$ and the similar between NT and PT. On the contrary, when $\kappa>1$, the QL weights are lower in NT than PT for almost the full ion scale. These observations suggest that the nonlinear heat fluxes will follow the same trend that we expect, i.e. the beneficial effect of NT should be lost both in ITG and TEM regimes when $\kappa<0.75$.

\subsection{Nonlinear simulations}

\begin{figure}[htbp]
    \centering
    \includegraphics[width=0.85\textwidth]{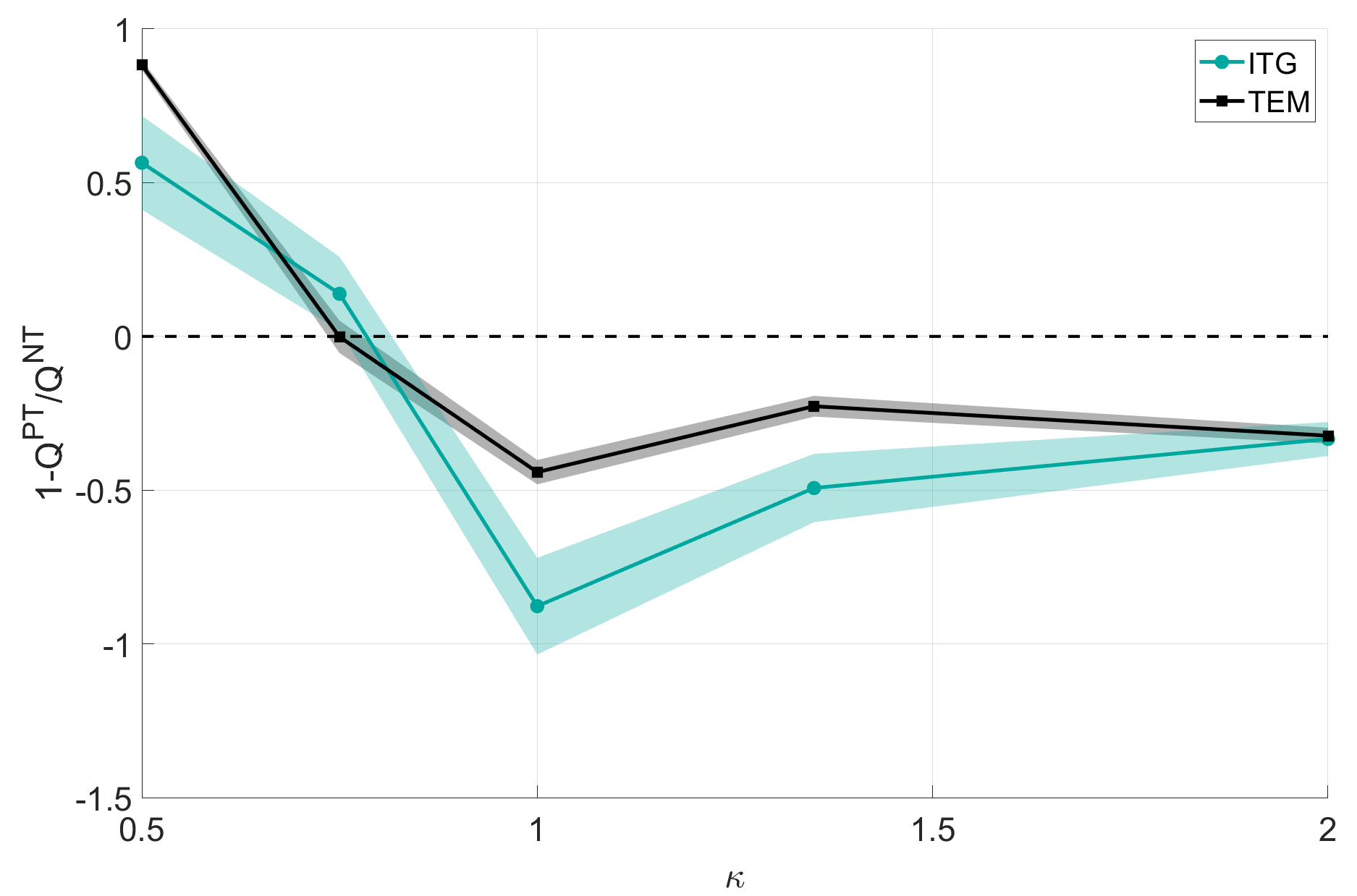}
    \caption{Relative difference of the nonlinear heat fluxes of PT and NT as a function of elongation $\kappa$ for the TEM (black) and ITG (green) cases.}
    \label{NL_scankappa}
\end{figure}

Using the same ITG and TEM cases, we performed an elongation scan with nonlinear simulations. Figure \ref{NL_scankappa} shows the ratio of nonlinear heat fluxes $1-Q^{PT}/Q^{NT}$. We observe that the general trend is in agreement with the results of our linear simulations and QL estimates, as well as with the predictions made by the geometric coefficients. Nonlinear simulations also show that NT has worse transport than PT for $\kappa<0.75$ and better (lower) transport for $\kappa>0.75$. This applies to both TEM- and ITG-dominated regimes. It is worth mentioning that the nonlinear simulations show a minimum in the relative difference between NT and PT around $\kappa=1.0$. This behaviour is not a surprise if we look again at the geometric coefficients shown in figure \ref{metrics_kappa}. We already pointed out that, while the relative difference of $\langle v_{Dy}\rangle$ and $\langle g^{yy}\rangle$ decreased monotonically, the relative difference of $\langle |dv_{Dy}/dz|\rangle$ has a minimum around $\kappa=1.0$. This trend is followed by the curves in figure \ref{NL_scankappa}, indicating the importance of $\langle |dv_{Dy}/dz|\rangle$. Comparing the ITG and TEM cases, it seems that the ITG scenario is more sensitive to the poloidal variation of the binormal magnetic drift velocity.

\begin{figure}[htbp]
    \centering
    \includegraphics[width=\textwidth]{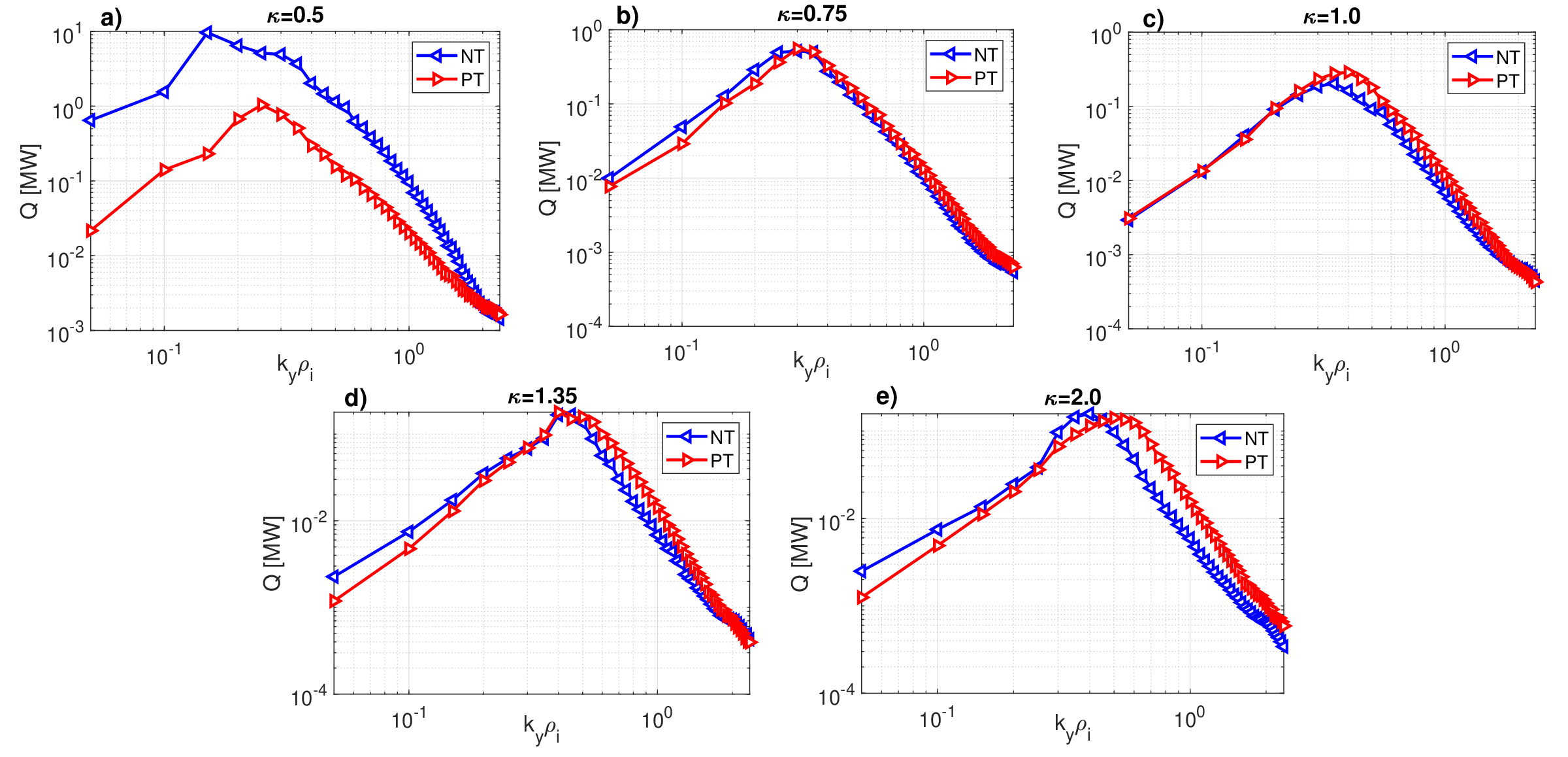}
    \caption{Total heat flux spectra for NT (blue) and PT (red) TEM-dominated cases at different values of elongation.}
    \label{Q_spectra_TEM}
\end{figure}

\begin{figure}[htbp]
    \centering
    \includegraphics[width=\textwidth]{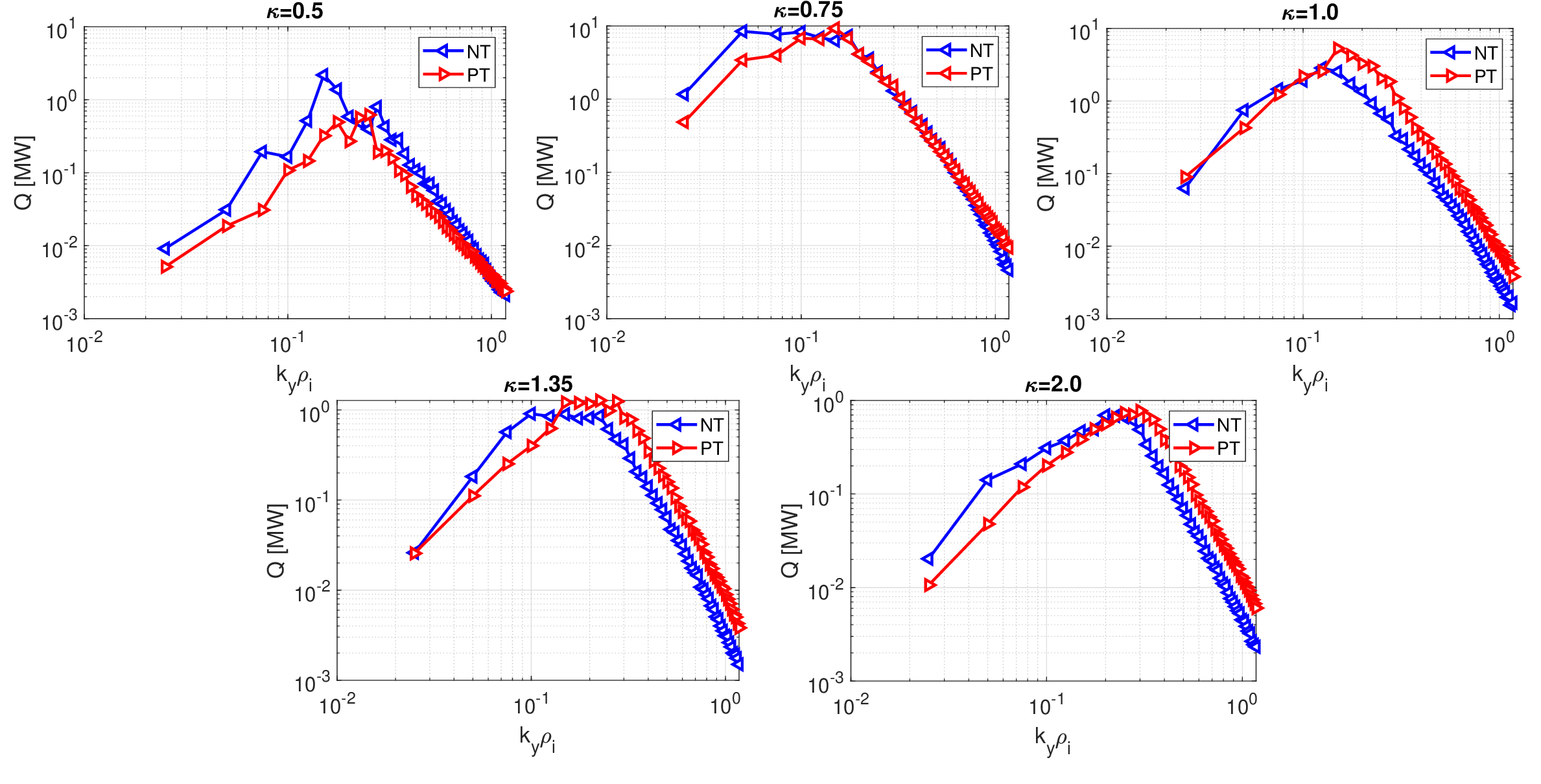}
    \caption{Total heat flux spectra for NT (blue) and PT (red) ITG-dominated cases at different values of elongation.}
    \label{Q_spectra_ITG}
\end{figure}

To gain more insight into the behavior of the turbulence, figures \ref{Q_spectra_TEM} and \ref{Q_spectra_ITG} show the spectra of the heat fluxes for the TEM and ITG cases respectively. When NT has a beneficial effect on turbulence, i.e. $\kappa>0.75$, we see that the stabilization involves the modes with $k_y\rho_i\gtrsim  0.3$. The modes with lower wavenumber are actually slightly more unstable than PT. However, since the modes with $0.3<k_y\rho_i<0.5$ contribute to most of the total heat flux, the overall effect of NT is stabilizing. These results are in qualitative agreement with the spectra predicted by the QL weights shown in figures \ref{QL_TEM_scankappa} and \ref{QL_ITG_scankappa}. On the other hand, when NT leads to larger transport than PT at low elongation, the modes at all wavenumbers are destabilized with respect to PT.

In conclusion, GK simulations confirm the predictions made by studying trends in the geometric coefficients. This provides additional evidence that faster magnetic drifts (which vary more along the poloidal cross-section) and stronger FLR effects can explain most of the beneficial effect of NT on ITG and TEM turbulence. To further validate this picture we performed experiments on TCV.

\section{Experimental elongation scan with negative and positive triangularity in TCV}\label{4}
% \subsection{Experimental observations}

\begin{figure}[htbp]
    \centering
    \includegraphics[width=\textwidth]{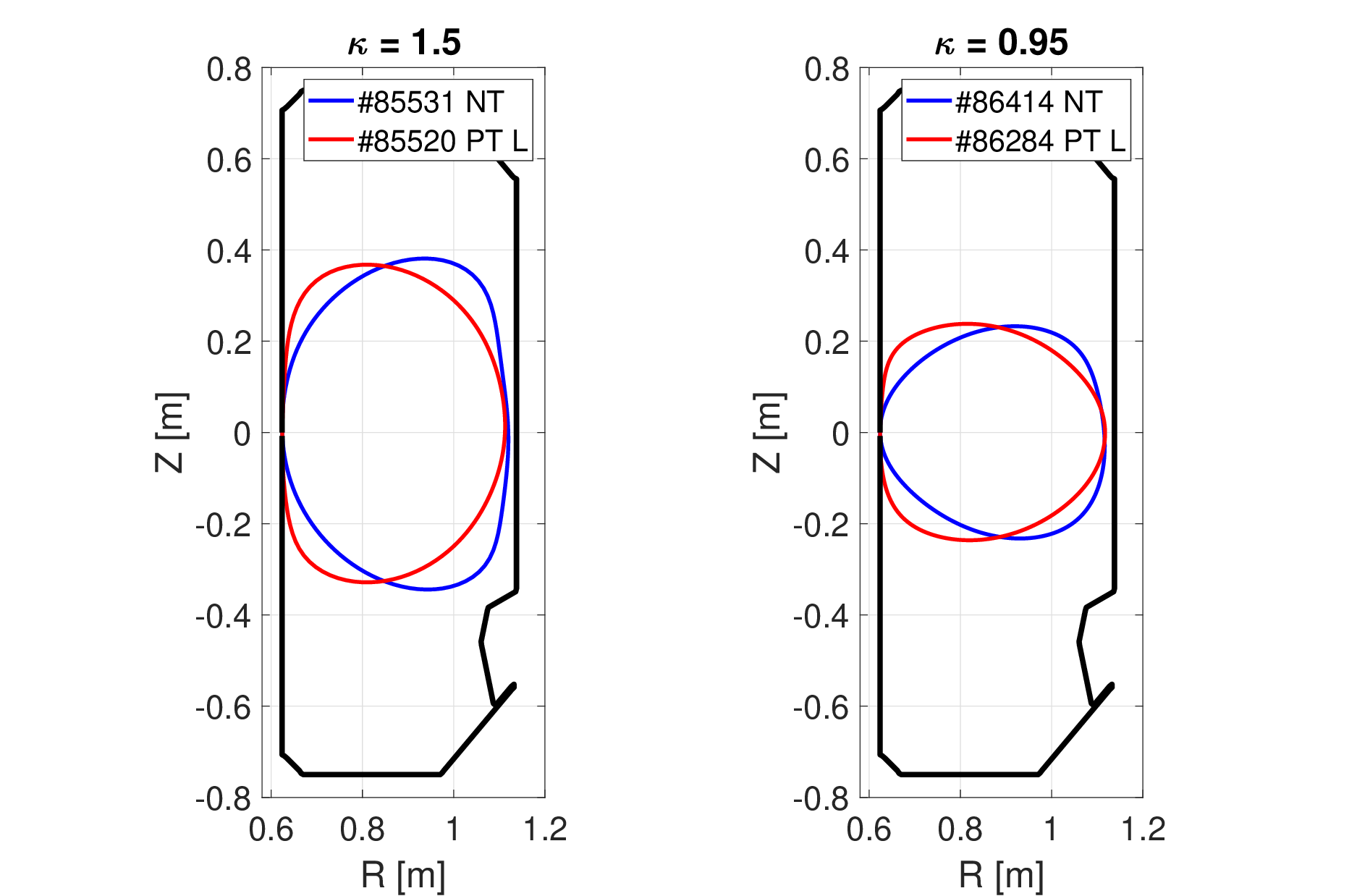}
    \caption{Shapes of the poloidal cross-section of the LCFS for the considered discharges.}
    \label{shapes_kappa}
\end{figure}

\begin{figure}[htbp]
    \centering
    \includegraphics[width=0.9\textwidth]{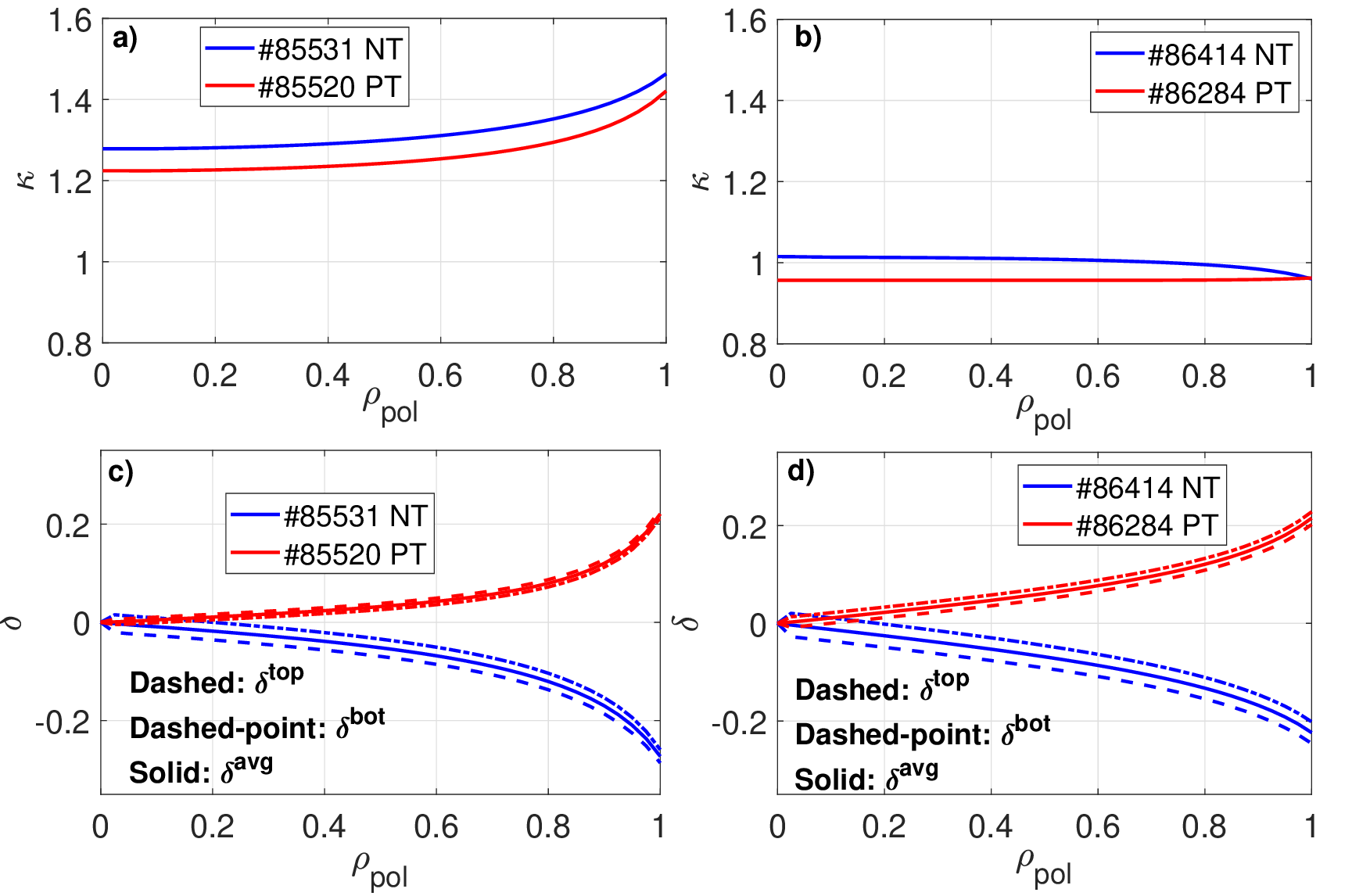}
    \caption{Elongation (top) and triangularity (bottom) of the considered discharges as functions of the radial coordinate $\rho_{pol}.$}
    \label{kappa_delta_vs_rho}
\end{figure}

In this section, we present the results of an elongation scan performed on TCV. Exploiting the geometric flexibility of TCV, we produced NT and PT inner-wall-limited discharges with moderately large Last Closed Flux Surface (LCFS) elongation $\kappa^{LCFS}=[1.4-1.5]$ and low LCFS elongation $\kappa^{LCFS}=0.95$. The LCFS of the achieved shapes are shown in figure \ref{shapes_kappa}, while the radial penetration of elongation and triangularity is shown in figure \ref{kappa_delta_vs_rho}. Figure \ref{kappa_delta_vs_rho} confirms that the magnitude of triangularity has been kept the same and only elongation has been changed between the two pairs of discharges. Smaller values of elongation were attempted. However, at elongations significantly smaller than unity, the plasma column became strongly radially unstable and difficult to control. 

We opted for limited discharges because they are easier to control and the NT geometry is less prone to fast Vertical Displacement Events (VDE), which can be problematic at high elongation. We do not expect diverted scenarios to behave differently. Nonetheless, such an investigation will be performed in the near future. The scenarios were heated with two levels of Electron Cyclotron Resonance Heating (ECRH), i.e. $P_{EC}=0.5$MW and $P_{EC}=1.0$MW. In this way it was possible to achieve a regime largely dominated by TEMs. In TCV, TEM-dominated regimes are more common than ITG-dominated regimes, especially in the outer core. For this reason, we decided to destabilize even further the TEMs instead of creating a mixed ITG-TEM scenario by using Neutral Beam Injection (NBI).

\begin{figure}[htbp]
    \centering
    \includegraphics[width=\textwidth]{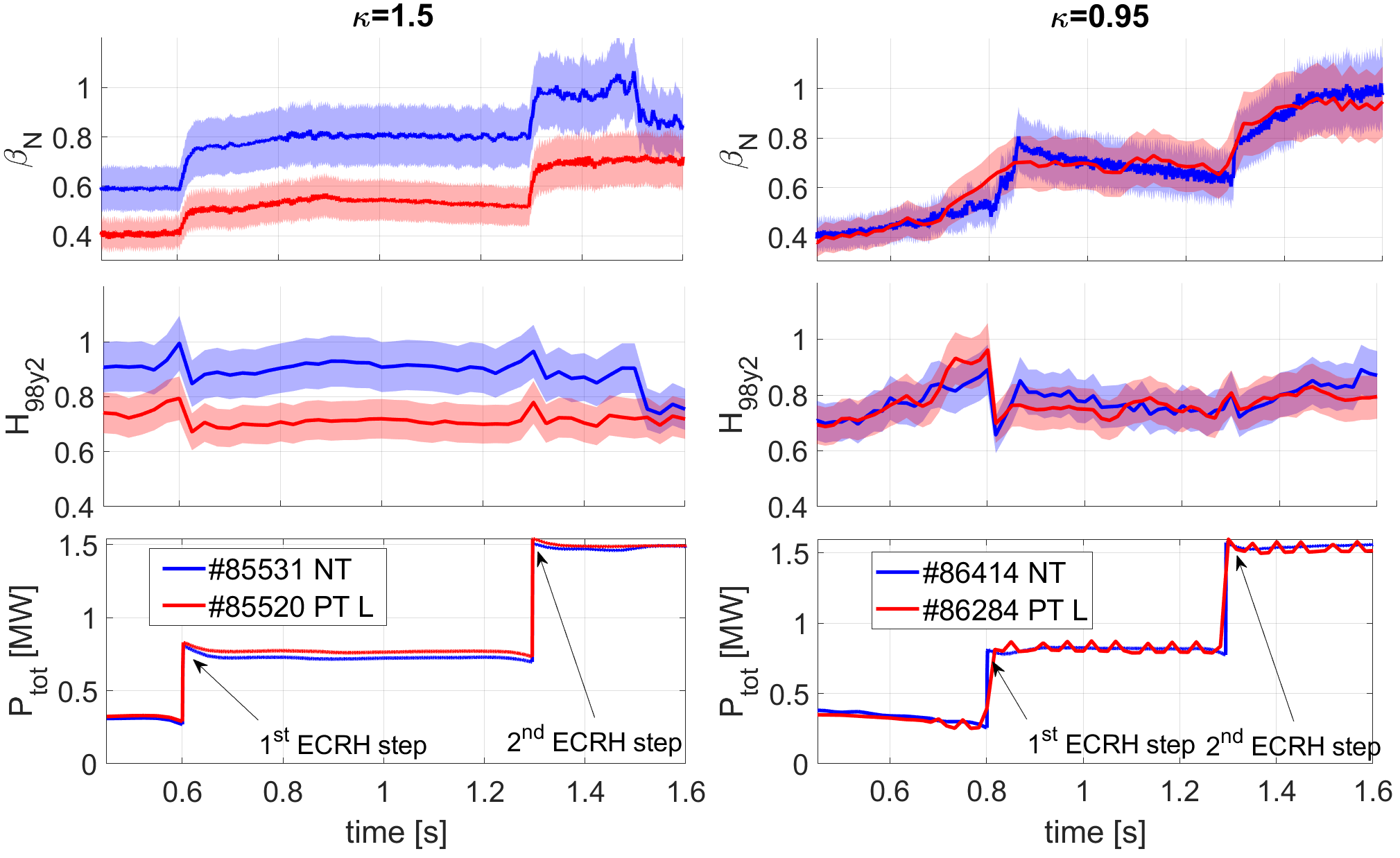}
    \caption{Time traces of $\beta_N$, $H_{98y2}$ and total input power in NT (blue) and PT (red) discharges at high elongation (left) and low elongation (right).}
    \label{time_traces_kappa}
\end{figure}

The comparison between NT and PT at a given elongation was made at fixed input power and the same fueling conditions. Therefore, we let the density and the temperature profiles evolve freely depending on the turbulent transport. We assess the performance and the quality of confinement of a scenario by looking at 0D quantities, i.e. $\beta_N$ and $H_{98y2}$, as well as at the plasma kinetic profiles, i.e. $T_e$, $n_e$ and $T_i$. The 0D parameters have been computed using the measurements of temperature and density profiles for ions and electrons coming from the TCV Thomson Scattering (TS) \cite{Blanchard_2019} and Charge eXchange Recombination Spectroscopy (CXRS) systems \cite{Bagnato_thesis}. Figure \ref{time_traces_kappa} shows the time traces of $\beta_N$, $H_{98y2}$ and the total input power $P_{tot}$. Here, the total input power includes the Ohmic contribution and ECRH. For the discharges with $\kappa=1.5$, we have a pure Ohmic phase between $t=[0.5,0.6]s$, then apply $0.5MW$ of ECRH between $t=[0.6,1.3]s$ and finally $1$MW of ECRH power after $1.3s$. For the $\kappa=0.95$ case, the ECRH steps are almost the same except that the $0.5MW$ injection starts slightly later at $t=0.8s$. In all the cases, we checked with the ray-tracing code TORAY \cite{Prater_2008} that all the injected power was absorbed well.

\begin{figure}[htbp]
    \centering
    \includegraphics[width=\textwidth]{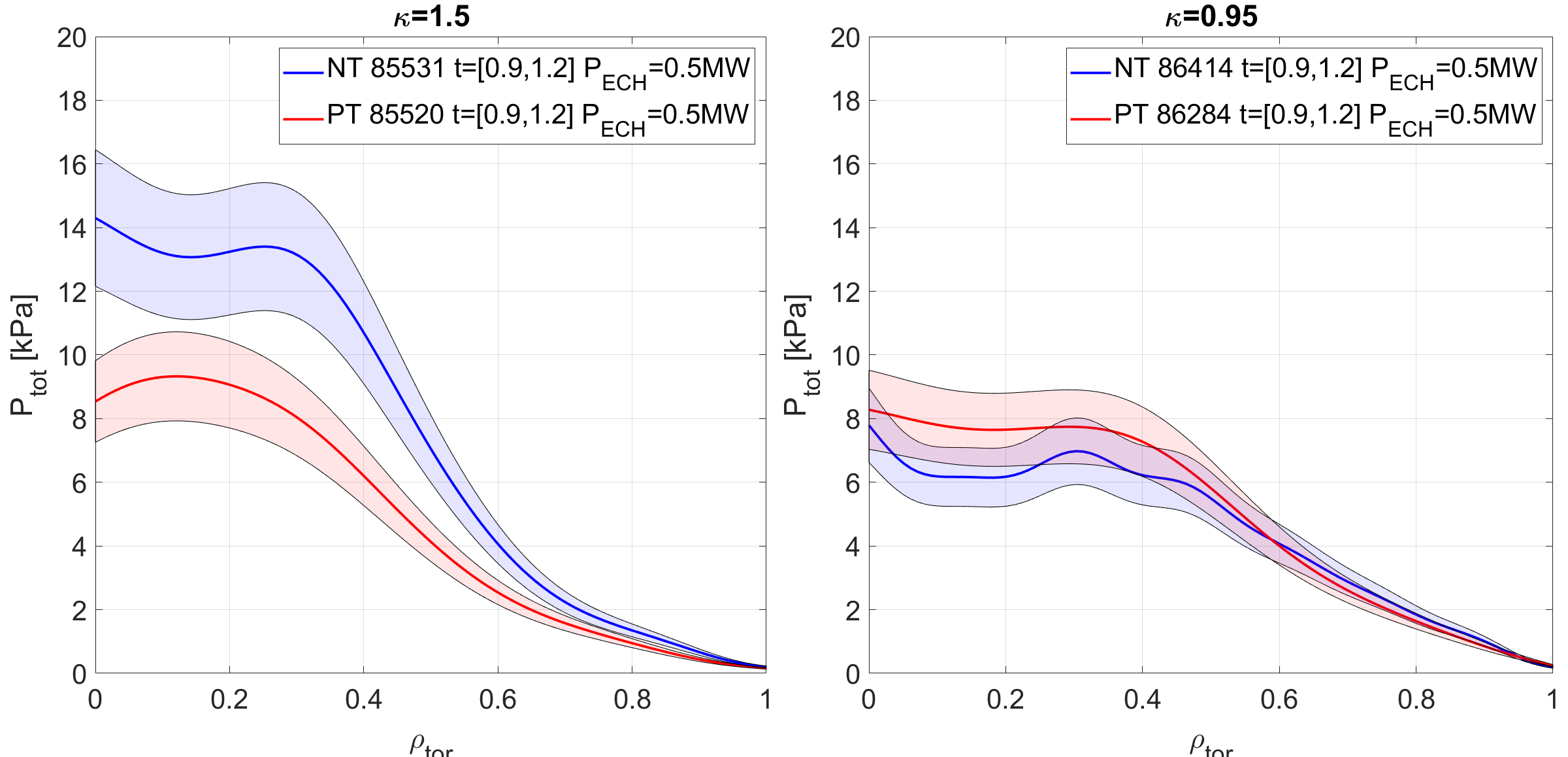}
    \caption{Radial profile of the total plasma pressure in NT (blue) and PT (red) at high (left) and low (right) elongation.}
    \label{pressure_scan_kappa}
\end{figure}

\begin{figure}[htbp]
    \centering
    \begin{subfigure}
        {\includegraphics[width=0.48\textwidth]{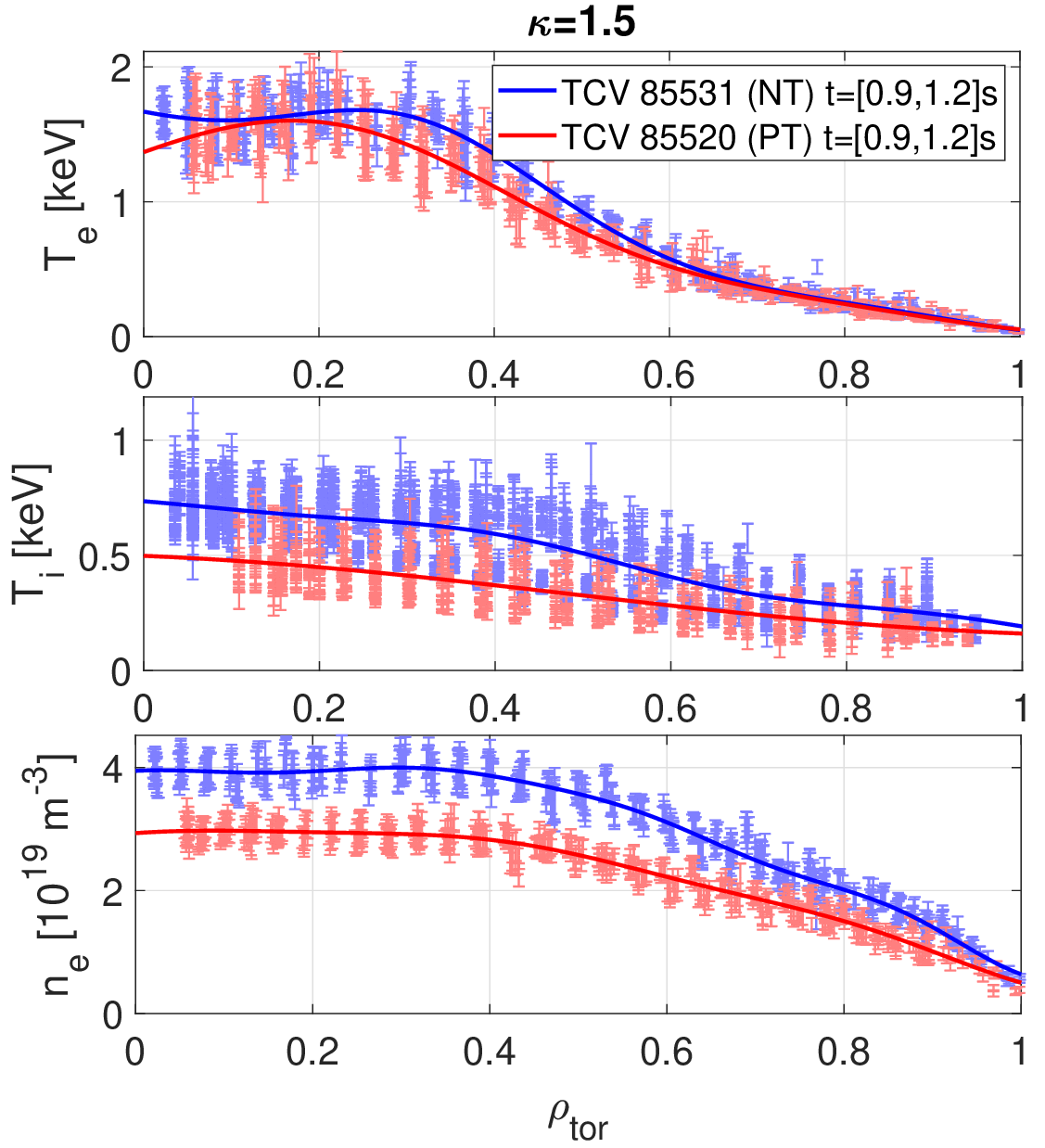}}
    \end{subfigure}
     \begin{subfigure}
        {\includegraphics[width=0.48\textwidth]{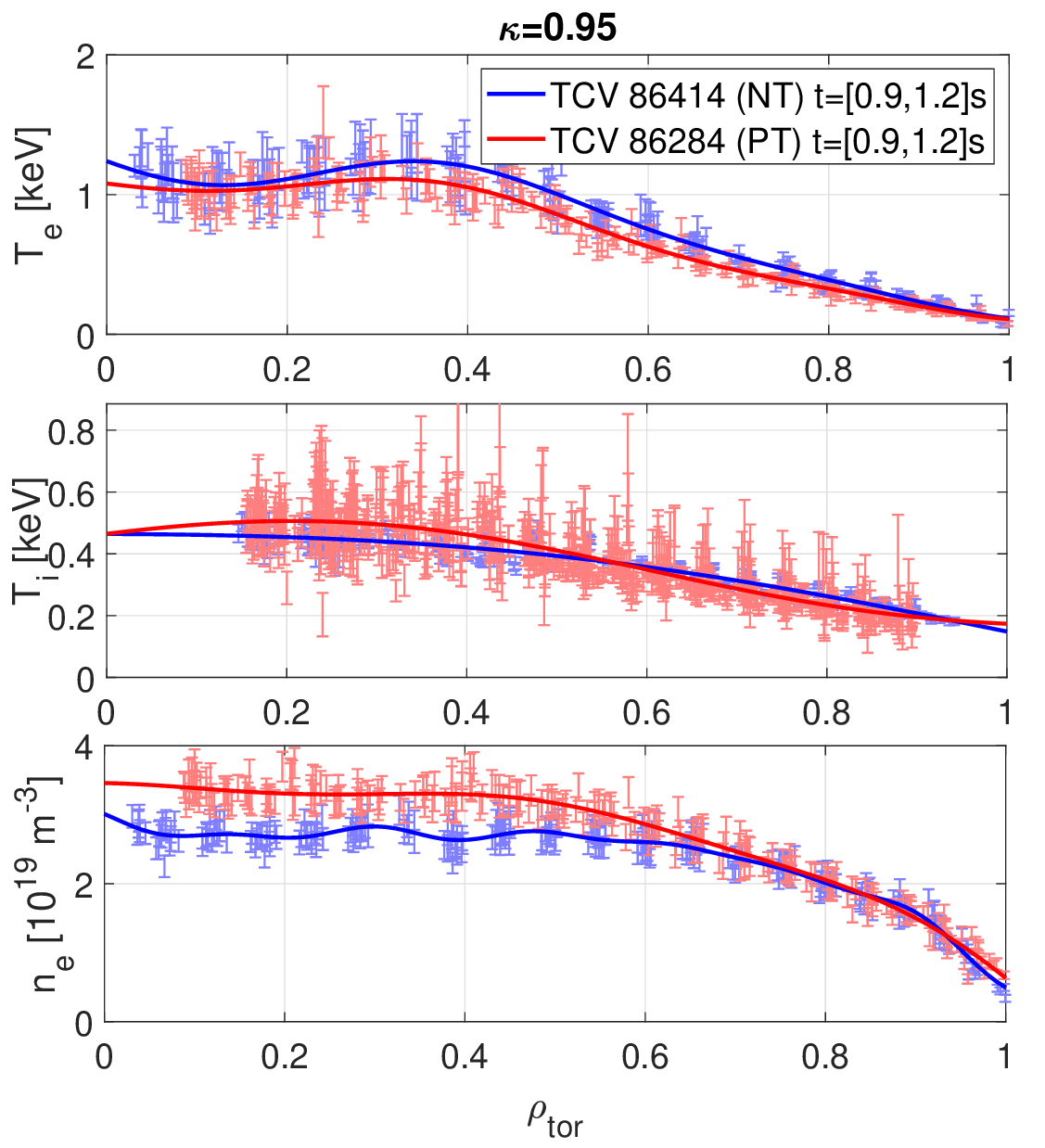}}
    \end{subfigure}
    \caption{Radial profiles of electron temperature (top), ion temperature (middle) and electron density (bottom) in NT (blue) and PT (red) at high (left) and low (right) elongation.}
    \label{profiles_scan_kappa}
\end{figure}

The left column of figure \ref{time_traces_kappa} shows the time traces for the high elongation case. We clearly observe that the NT case achieves better performance than the PT case. At the same input power, the NT shape reaches values of $\beta_N$ and $H_{98y2}$ that are 1.5 times larger than PT. In addition, the value of $H_{98y2}$ for NT is close to unity for the whole discharge, showing H-mode-like confinement. Finally, we note that the sudden decrease in performance of the NT case at $t=1.5s$ is due to a Neoclassical Tearing Mode (NTM) that eventually leads to a fast VDE and a disruption. The right column of figure \ref{time_traces_kappa} shows the time traces of the low $\kappa$ scenarios. Contrary to the high elongation cases, there is now little difference between NT and PT. Both the traces of $\beta_N$ and H-factor are similar. To confirm these trends, we can also look at the kinetic profiles. In figure \ref{pressure_scan_kappa} we show the radial profile of the total pressure for the two elongations, while in figure \ref{profiles_scan_kappa} we display the radial profiles of $T_e$, $n_e$ and $T_i$ for all the considered discharges. The radial profiles are computed in the time window $t=[0.9,1.2]s$, i.e. with $P_{EC}=0.5$MW. The same trend holds for the second power step, but it is not shown here for brevity.

Figure \ref{pressure_scan_kappa} shows that the improved performance in NT at $\kappa=1.5$ comes from an increase in total plasma pressure across the whole radial domain. On the other hand, at $\kappa=0.95$, we see that the total pressure profile of the NT and PT cases almost overlap, confirming the observation that the beneficial effect of NT on turbulence degrades at low elongation. Looking at the kinetic profiles in figure \ref{profiles_scan_kappa}, we see that the improvement in confinement for the NT case at $\kappa=1.5$ comes from larger values of $T_i$ and $n_e$. This is consistent with observations made in \cite{Balestri_2024} for a DTT-NT similarity experiment performed on TCV with only ECRH heating. On the other hand, at $\kappa=0.95$ the NT and PT cases have similar profiles of $T_e$, $T_i$ and $n_e$.

\begin{figure}[htbp]
    \centering
    \includegraphics[width=\textwidth]{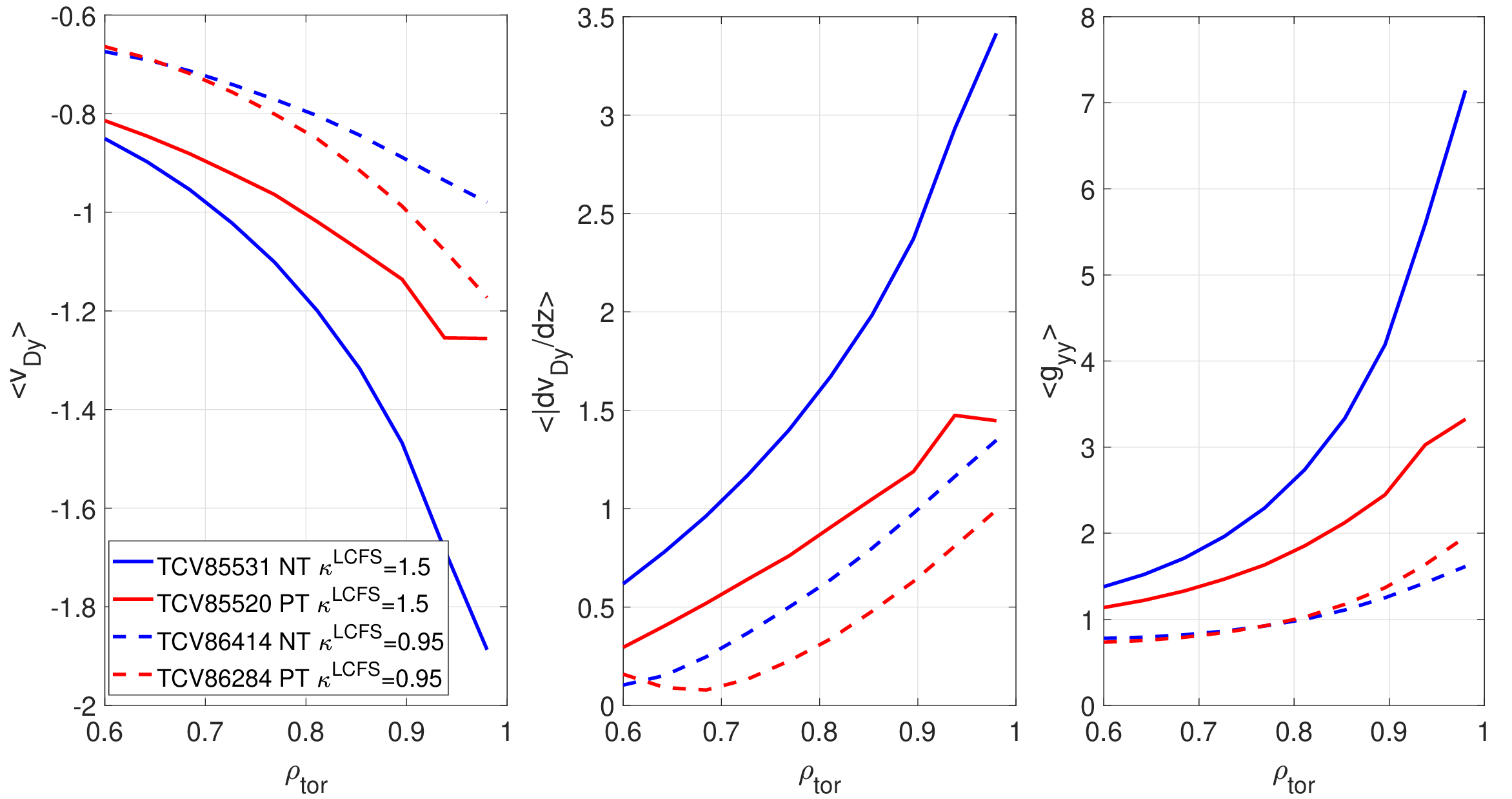}
    \caption{ The poloidally averaged (left) magnetic drift $\langle v_{Dy}\rangle$, (center) absolute value of the variation of the magnetic drift $\langle |dv_{Dy}/dz|\rangle$ and (right) FLR effects $\langle g^{yy}\rangle$ in NT (blue) and PT (red) geometry at high (solid) and low (dashed) elongation as functions of the radial coordinate $\rho_{tor}$. The poloidal averages have been performed around the outboard midplane, i.e. $\theta=[-1,1]$.}
    \label{metrics_exp}
\end{figure}

As a final check, we compute the geometric proxies at different radial locations for the experimental equilibria. Figure \ref{metrics_exp} shows that the magnetic drift and FLR metrics behave as expected. For the high elongation cases (solid lines), NT exhibits faster magnetic drifts and stronger FLR effects than PT, consistent with the better confinement performance observed experimentally. For the low elongation cases (dashed lines), the two proxies flip, with PT showing somewhat larger values. This is in agreement with the absence of any measurable difference in performance between the two configurations at low elongation.

The only metric that does not neatly follow this trend is the variation of the magnetic drift $\langle|dv_{Dy}/dz|\rangle$, which remains larger in NT than in PT at both elongations. Two considerations help to understand this. First, the gap between $\langle|dv_{Dy}/dz|\rangle$ in NT and PT narrows considerably as elongation decreases, so this coefficient should be less beneficial for NT at $\kappa=0.95$ compared to $\kappa=1.5$. Thus, it seems reasonable that the weaker stabilizing effect of $\langle|dv_{Dy}/dz|\rangle$ on NT is outweighed by the destabilizing effect of $\langle v_{Dy}\rangle$ and $\langle g^{yy}\rangle$. Second, as shown in figure \ref{metrics_kappa}, at $\kappa=0.95$ the relative difference of the coefficients has not fully flipped signs. Indeed, $\langle|dv_{Dy}/dz|\rangle$ is still predicted to be stabilizing. Ideally, we would also perform experiments with $\kappa=0.75$, but such shapes have proved challenging to achieve stably. We conclude that, overall, the geometric analysis of the experimental equilibria provides a satisfactory and consistent validation of the physical picture developed in this work.

In summary, analysis of the geometric coefficients indicates that reducing elongation below $\kappa\sim1$ leads to a degradation of the beneficial effect of NT. For $\kappa\lesssim0.8$, we expect NT to have worse confinement than PT. TCV experiments at $\kappa=1.5$ and $\kappa=0.95$ show behaviour consistent with this prediction.

% We conclude that the analysis of the 0D quantities and the kinetic profiles confirms and validate our predictions. In a TEM-dominated plasma a reduction of elongation below unity leads to a degradation of the beneficial effect of NT and, eventually, to worse performance in NT than PT if elongation is sufficiently low. However, we have to point out that the metric coefficients and the GK simulations predicted the reversal of the beneficial effect of NT around $\kappa=0.75$. Nonetheless, it is important to keep in mind that the magnetic equilibria produced in the experiments are different from the ones used to perform the simulations. Hence, we were not expecting a perfect match, but to reproduce the expected trend.% Therefore, to further validate our picture, we used the new experimental equilibria at $\kappa=1.5$ and $\kappa=0.95$ to compute the metric coefficients.

\section{Verification in realistic ITG scenarios across multiple tokamaks}\label{5}

In this section we apply our physical picture to realistic ITG-dominated scenarios taken from DIII-D, AUG and JET, spanning a range of plasma shapes, including up-down asymmetric configurations. In doing so, we will analyze interesting experimental results and disentangle the role of magnetic drifts and FLR effects from that of parallel dynamics. So far, we focused only on the predictive capabilities of the magnetic drifts and FLR effects, neglecting the potential role of parallel dynamics, i.e. the parallel streaming and mirror terms. We will first present the equilibria considered and then the simulation results.

%This provides a broad validation of the metric-based framework introduced in chapter~\ref{ch5}.

\subsection{Experimental scenarios and equilibrium parameters}

\subsubsection*{DIII-D \#171421}
We consider a high-confinement limited NT discharge from DIII-D (\#171421 at $t=2.8\,\text{s}$), taken from \cite{Marinoni_2024}. This is a well diagnosed limited NT plasma that showed good confinement properties, reaching $H_{98y2}=0.95$. Some GK simulations of this discharge have already been carried out in \cite{Marinoni_2024}, showing a turbulent regime where ITG modes are unstable and are responsible for a big fraction of the total transport. Using GK linear and nonlinear simulations, we will compare the transport of the NT case with an artificial PT scenario produced by flipping the boundary of the reference NT case. We then swap the geometric coefficients responsible for magnetic drifts, FLR effects and parallel dynamics to estimate their impact. 
\begin{figure}[htbp]
    \centering
    \includegraphics[width=0.45\textwidth]{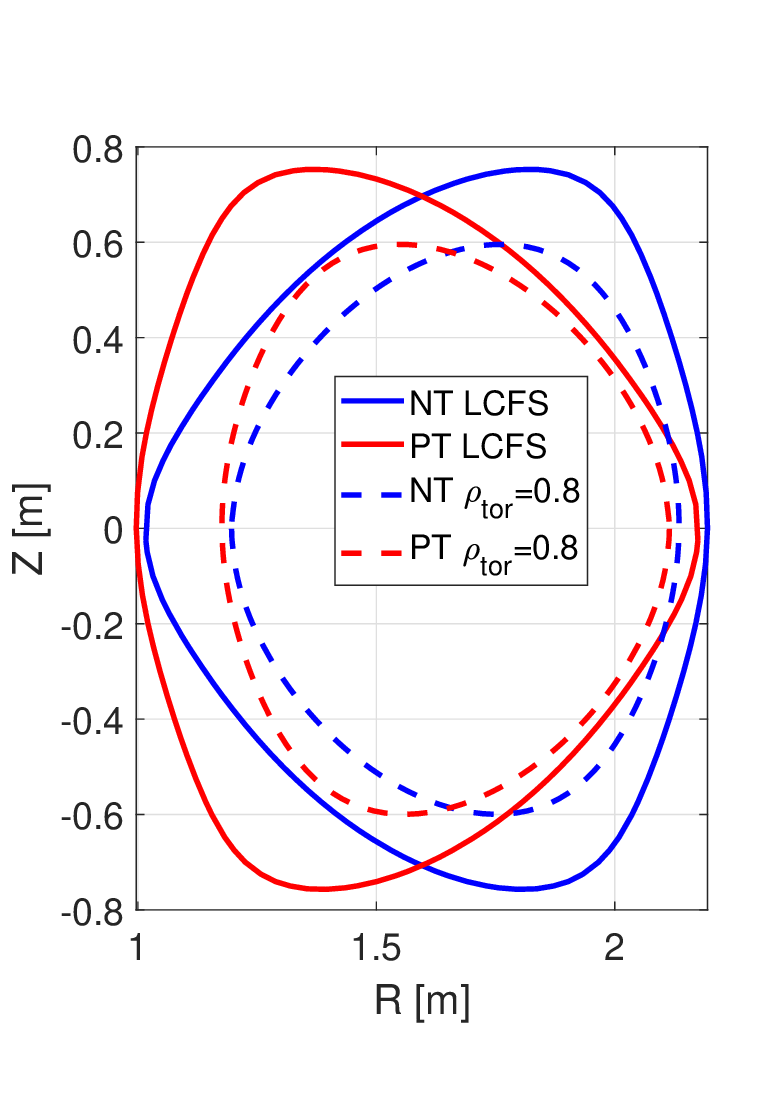}
    \caption{The poloidal cross-section of the LCFS and simulated flux surfaces from the NT and PT scenarios for DIII-D \#171421.}
    \label{DIIID_eqdsk}
\end{figure}
\begin{figure}[htbp]
    \centering
    \includegraphics[width=1\textwidth]{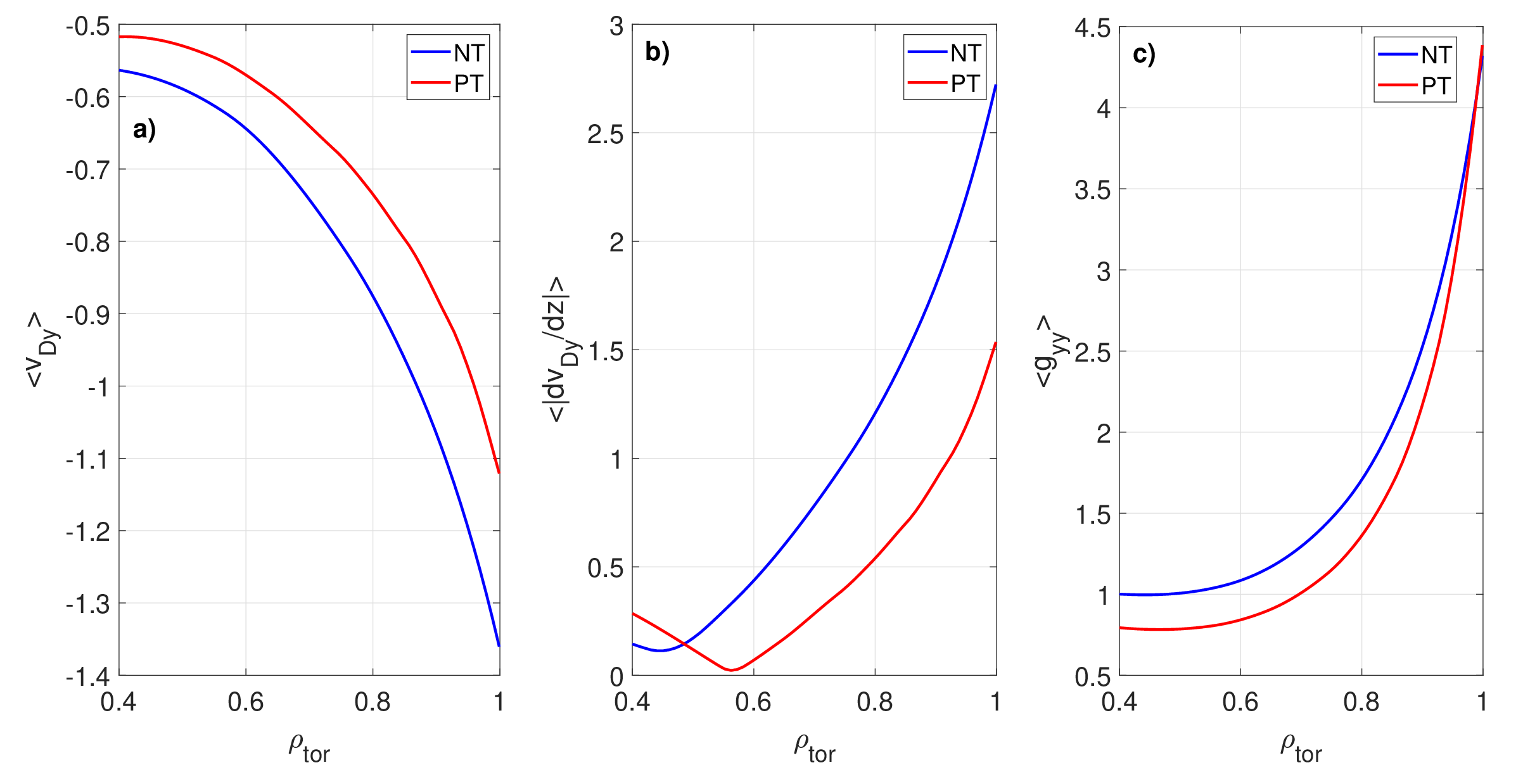}
    \caption{The poloidally averaged (a) magnetic drift $\langle v_{Dy}\rangle$, (b) absolute value of the variation of the magnetic drift $\langle |dv_{Dy}/dz|\rangle$ and (c) FLR effects $\langle g^{yy}\rangle$ in NT and PT geometry as functions of the radial coordinate $\rho_{tor}$ for DIII-D. The poloidal averages have been performed around the outboard midplane, i.e. $\theta=[- 1,1]$.}
    \label{DIIID_metrics}
\end{figure}
The LCFS and the simulated flux surfaces of the NT and PT scenarios are shown in figure \ref{DIIID_eqdsk} while the poloidally averaged geometrical proxies, i.e. $\langle v_{Dy}\rangle$, $\langle|dv_{Dy}/dz|\rangle$ and $\langle g^{yy}\rangle$ for NT and PT are shown in figure~\ref{DIIID_metrics} as functions of radius. We see that across the full radial domain the NT geometry displays more favourable values for all three quantities: a larger averaged magnetic drift, more variation in the drift along the poloidal angle, and stronger FLR effects. These observations lead us to predict a reduction of ITG turbulence in NT. 

\subsubsection*{AUG \#43493}
For AUG we consider discharge \#43493 from the most recent NT campaign, which achieved the strongest diverted NT shaping  yet on this machine and is marginal to H-mode access \cite{Vanovac_2026}. The scenario features $P_{tot}=10\,\text{MW}$ of NBI heating and $H_{98y2}=0.7$. We see that the H-factor is lower than unity. This suggests that the shape and scenario developed on AUG are still not optimal to achieve the reduction of transport observed on TCV and DIII-D. However, we must specify that this scenario has not been compared with corresponding PT L-mode and PT H-mode plasmas. As highlighted in \cite{Balestri_2024}, the H-factor is a volume averaged quantity that might penalize NT. Indeed, the lack of a pedestal, which develops at the edge of the plasma where volume is the largest, will tend to make the H-factor in NT lower than in a H-mode, even though the NT plasma is capable of reaching the same core pressure of the corresponding PT H-mode.

\begin{figure}[htbp]
    \centering
    \includegraphics[width=0.45\textwidth]{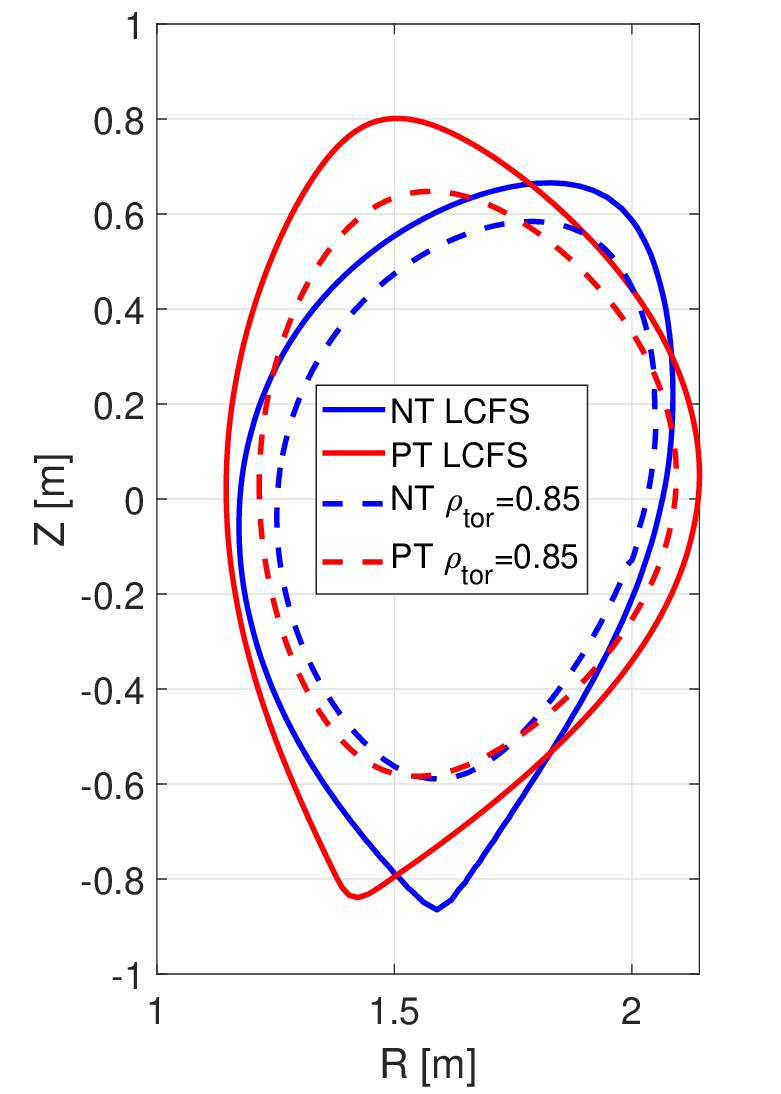}
    \caption{The poloidal cross-section of the LCFS and simulated flux surfaces from the NT and PT scenarios for AUG43493 and AUG36157.}
    \label{AUG_eqdsk}
\end{figure}

\begin{figure}[htbp]
    \centering
    \includegraphics[width=\textwidth]{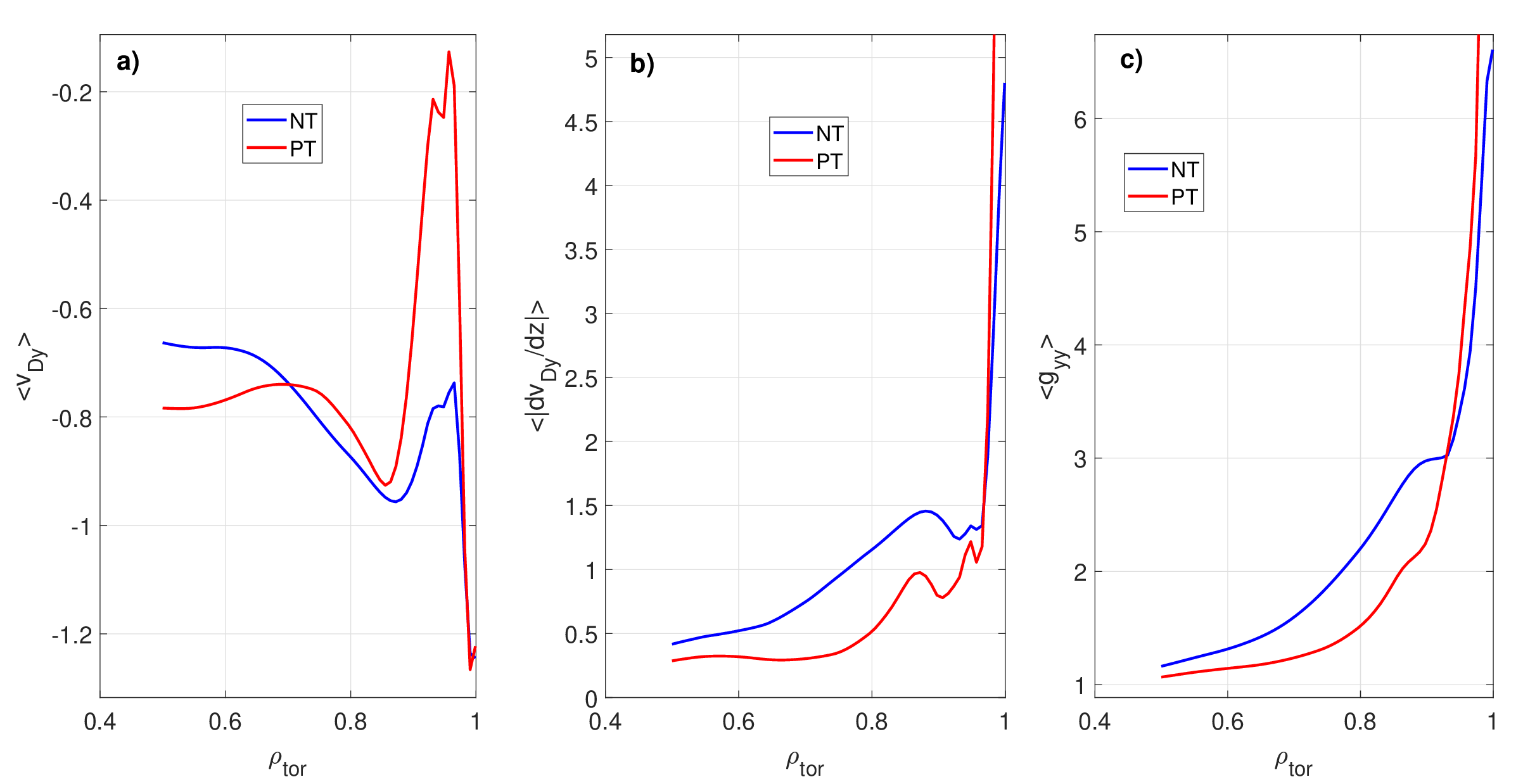}
    \caption{The poloidally averaged (a) magnetic drift $\langle v_D\rangle$, (b) absolute value of the variation of the magnetic drift $\langle |dv_D/dz|\rangle$ and (c) FLR effects $\langle g^{yy}\rangle$ in NT and PT geometry as functions of the radial coordinate $\rho_{tor}$ for AUG. The poloidal averages have been performed around the outboard midplane, i.e. $\theta=[-1,1]$.}
    \label{AUG_metrics}
\end{figure}

With simulations, we compare the AUG NT scenario with an AUG PT discharge, shot \#36157 taken from \cite{Aucone_2024}, where we consider only the shape and run the simulations using the profiles of the NT scenario. Discharge \#36157 is an ELM-free H-mode with a clear LH transition at $800\; \text{kA}$. On the other hand, the NT discharge has 600 kA of plasma current, which will result in a higher $q_{95}$ and in general a different safety factor profile. The fact that the difference between the two scenarios goes beyond simple triangularity is captured by the geometric proxies, highlighting once more their usefulness. The comparison between the two different shapes at fixed gradients allows us to have an estimate of how much NT gains in terms of reduction of transport with respect to the PT scenario. The LCFS and the simulated flux surfaces of the NT and PT scenarios are shown in figure \ref{AUG_eqdsk}.  Then, we will test the reliability of our geometrical proxies in predicting the actual difference between NT and PT. As we can see from figure \ref{AUG_metrics}, the NT and PT shapes are characterized by different coefficients. In figure \ref{AUG_metrics}(a) we see that the average magnetic drift is more negative in NT than PT for $\rho_{tor}\geq0.7$. From figure \ref{AUG_metrics}(b) we see that the drift variation is larger in NT than in PT across the considered radial domain, except for a narrow region with $\rho_{tor}>0.97$ and from figure \ref{AUG_metrics}(c) we observe that the FLR effects are larger in NT than PT for $\rho_{tor}<0.95$. Therefore, according to our model, we expect to see a significant reduction of transport in the range $0.7<\rho_{tor}<0.95$, where \textit{all} three geometrical proxies indicate a benefit. As for DIII-D, we will test this hypothesis with linear and nonlinear simulations.

It is worthy to mention that all three geometrical proxies indicate that NT should be worse than PT for $\rho>0.97$. Our model has not been validated for radial locations so close to the LCFS, but this observation is compatible with the not excellent performance of the AUG NT scenario observed experimentally. Our proxies seem to indicate that the reduction of turbulent transport is significant in the outer core, but is then lost very close to the LCFS, thus, potentially inhibiting the considerable improved confinement observed on TCV and DIII-D. However, we stress once more that our model has not been validated so close to the LCFS and more work is needed to validate this hypothesis.

% \begin{table}[htbp]
%     \centering
%     \begin{tabular}{l|cc}
%     \toprule
%         &NT & PT  \\ \hline
%         $R/L_{Te}$&11.7 & 11.7 \\
%         $R/L_{Ti}$&9.11  & 9.11\\
%         $R/L_{ne}$&1.59  & 1.59\\
%         $T_e$ [keV]&0.58  & 0.58\\
%         $T_i/T_e$&1.03  & 1.03\\
%         $n_e\,[10^{19}m^{-3}]$& 3.72 & 3.72\\
%         $\beta [\%]$ &0.14  & 0.14\\
%         \hline
%         $A$& 4  & 4 \\
%         $q$&4.39 & 3.70\\
%         $\hat{s}$& 2.46 & 2.97\\\toprule
%     \end{tabular}
%     \caption{\label{AUG_input} Input parameters for AUG \#43493 at $\rho_{tor}=0.85$. The shape is parametrized via Fourier-Miller; individual geometric shaping coefficients beyond $q$ and $\hat{s}$ are not listed.}
% \end{table}

\newpage
\subsubsection*{JET \#105835} 
\begin{figure}[htbp]
    \centering
    \includegraphics[width=0.45\textwidth]{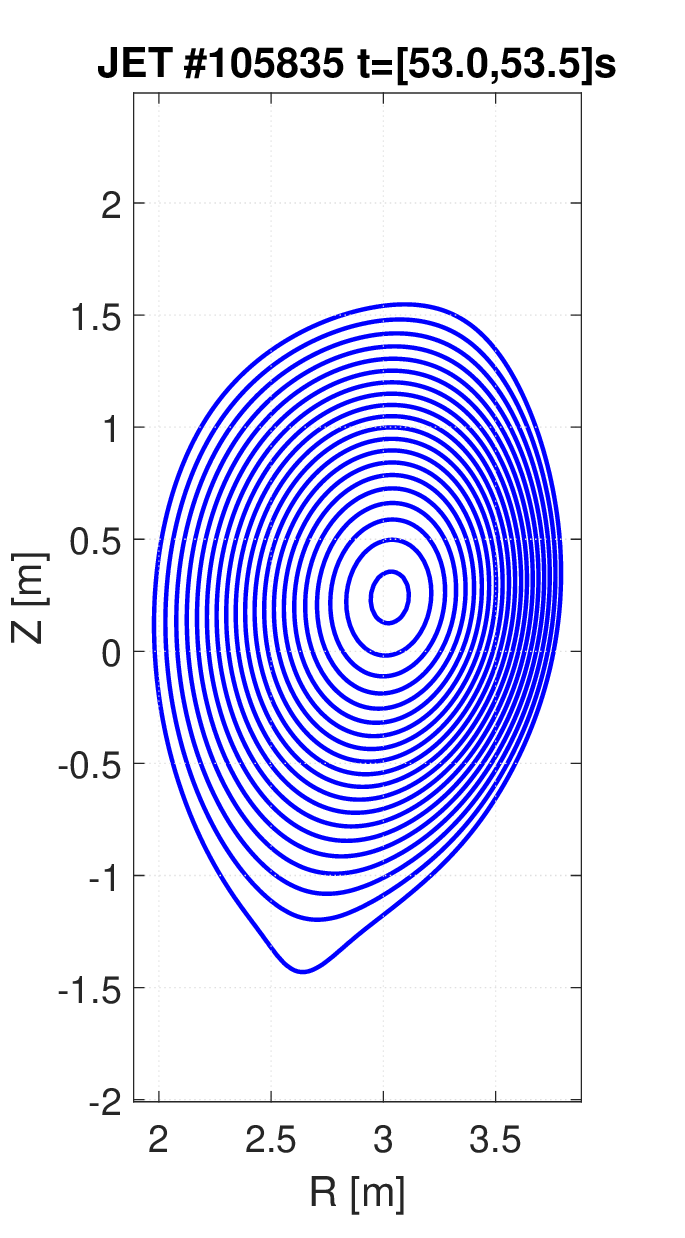}
    \caption{The poloidal cross-section of the LCFS and simulated flux surfaces from the NT scenarios of discharge JET105835.}
    \label{JET_eqdsk_single}
\end{figure}

The last scenario we consider to test our physical picture in realistic ITG-dominated scenarios is a NT discharge performed on JET (JPN 105835). More information on the experiment can be found in \cite{Sauter2023_APS,Dunne_2026} and here we provide only a short summary. As for AUG, the development of up-down strongly shaped NT plasmas in JET is constrained by the vacuum vessel, which is optimized to host PT plasmas. The shape achieved in JET is shown in figure \ref{JET_eqdsk_single}.
The scenario was developed and executed in a very short time, in only one afternoon session on JET during the last week of operation before the final shutdown of the machine. This is noteworthy as limited time was available to optimize the scenario. For the following analysis we chose the time window $t=[53,53.5]s$, where the plasma is heated with 20MW of NBI power and achieves $H_{98y2}=0.82$. The main target of the experiment was to demonstrate in a metallic and large machine like JET that H-mode access could be prevented with sufficiently strong shaping even at very large input power. Little attention was dedicated to confinement optimization given the short experimental time and the limited shaping capabilities of JET. Nonetheless, it is remarkable the good confinement that has been achieved.

Using GK linear and nonlinear simulations, we will compare the transport of the NT case with an artificial PT scenario produced by flipping and symmetrizing the boundary of the reference NT case.
%We were able to find a PT H-mode scenario sufficiently similar to this NT scenario. Specifically, we use discharge \#105754, which is a PT H-mode with an NBI input power of 20MW in the time interval $t=[44.5,45.5]s$. However, it has to be noted that the two discharges have slightly different plasma currents: $I_p=1.5MA$ for NT and $I_p=1.4MA$ for PT. 

%In figure \ref{JET_profiles} we compare the kinetic profiles and we can clearly see the beneficial effect of NT on confinement. Indeed, without the formation of the pedestal, NT is able to recover and even overcome the values of density, temperature and pressure of the H-mode in the core. We have to note that the steepening of $T_i$ in the NT case starting from $\rho_{tor}=0.4$ is probably not due to triangularity given that its absolute value is too low so deep in the core. In addition, there are indications that this might be due to NBI-generated fast ions.
% \begin{figure}[htbp]
%     \centering
%     \includegraphics[width=\textwidth]{images/JET_profiles_nice.eps}
%     \caption{Radial profiles of (a) electron density, (b) electron temperature, (c) ion temperature and (d) electron pressure for the JET NT (blue) and PT H-mode (red) discharges.}
%     \label{JET_profiles}
% \end{figure}
% \newpage
\subsection{Linear simulations}

The scenarios described in the previous section have been used to perform linear and nonlinear flux tube GK simulations. The input parameters used are shown in table \ref{DIIID_AUG_JET_input}.
\begin{table}[htbp]
    \centering
    \begin{tabular}{l|cc|cc|cc}
    % \hline
    &\multicolumn{2}{c}{DIII-D $\rho_{tor}=0.8$} & \multicolumn{2}{c}{AUG $\rho_{tor}=0.85$} & \multicolumn{2}{c}{JET $\rho_{tor}=0.85$}\\
    & \textcolor{blue}{NT} & \textcolor{red}{PT} & \textcolor{blue}{NT} & \textcolor{red}{PT} & \textcolor{blue}{NT} & \textcolor{red}{PT}\\\hline
        $R/L_{Te}$& 11.39 & 11.39 & 11.7 & 11.7 & 23.9 & 23.9   \\
        $R/L_{Ti}$& 9.27 & 9.27 & 9.1 & 9.1 & 19.7 & 19.7 \\
        $R/L_{ne}$& 3.72 & 3.72 & 1.6 & 1.6 & 5.12 & 5.12  \\
        $T_e$ [keV]& 0.46 & 0.46 & 0.58 & 0.58 & 0.75 & 0.75  \\
        $T_i/T_e$& 1.7  & 1.7 & 1.03 & 1.03 & 1.23 & 1.23 \\
        $n_e\,[10^{19}m^{-3}]$& 2.88 & 2.88 & 3.73 & 3.73 & 2.41 & 2.41 \\
        $\beta [\%]$ &0.13 & 0.13 & 0.14 & 0.14 & 0.13 & 0.13  \\
        \hline
        $A$& 4 & 4 & 3.8 & 3.8 & 4.5 & 4.5 \\
        $q$& 1.99 & 1.99 & 4.39 & 3.69 & 3.78 & 3.78 \\
        $\hat{s}$& 1.75 & 1.75 & 2.46 & 2.97 & 2.62 & 2.62 \\
        $\kappa$ & 1.28 & 1.28 & - & - & - & 1.5 \\
        $|\delta|$ &0.16 & 0.16 & - & - & - & 0.3 \\
        $\zeta$ & -0.05 & -0.05 & - & - & - & -0.001 \\
        $s_\kappa$ & 0.04 & 0.04 & - & - & - & 0.33\\
        $|s_\delta|$& 0.34 & 0.34 & - & - & - & 0.6  \\
        $s_\zeta$ &-0.13 & -0.13 & - & - & - & -0.1 \\
        $\partial_rR$ &-0.35 & -0.35 & - & - & - & -0.3 \\\hline
    \end{tabular}
    \caption{\label{DIIID_AUG_JET_input} Input parameters for DIII-D, AUG and JET, showing the logarithmic gradients of electron temperature $R/L_{Te}$, ion temperature $R/L_{Ti}$, electron density $R/L_{ne}$, ion-electron temperature ratio $T_i/T_e$, electron density $n_e$, local electron $\beta$, and Miller equilibrium geometric parameters.}
\end{table}
The DIII-D NT and PT magnetic equilibria have been parametrized using the local standard Miller parametrization because the flux surfaces are up-down symmetric, so this parametrization is sufficiently accurate. The AUG and JET NT equilibria have been parametrized with the Fourier version of the local Miller parametrization to fully capture the up-down asymmetry for the considered shapes.

\begin{figure}[htbp]
    \centering
    \includegraphics[width=0.9\textwidth]{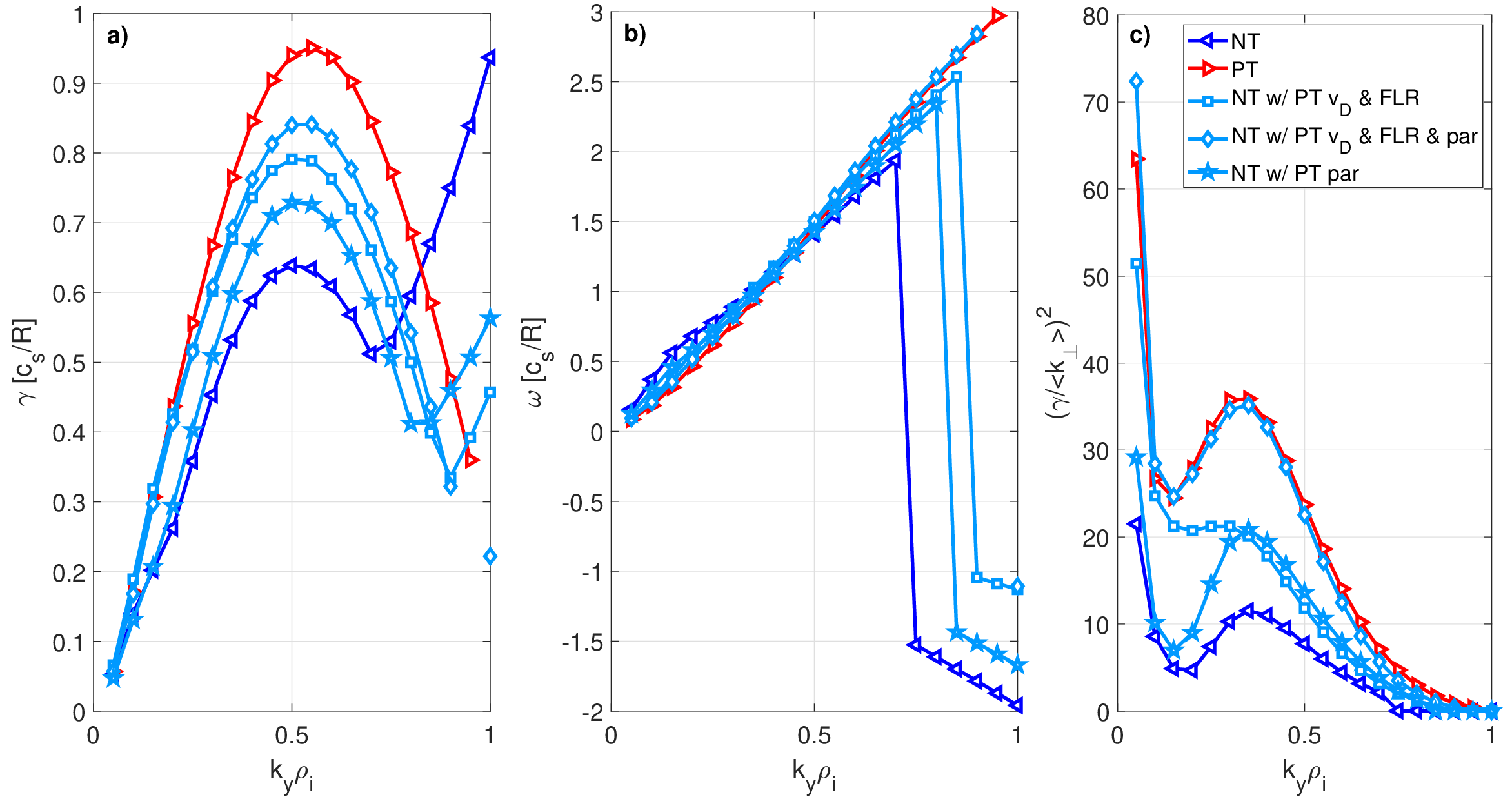}
    \caption{(a) Growth rates, (b) real frequencies and (c) QL weights as functions of $k_y\rho_i$ for NT (blue), PT (red) and cases with artificially swapped coefficients (light blue) for DIII-D.}
    \label{DIIID_gamma_omega_weight_vs_ky_swap}
\end{figure}

\begin{figure}[htbp]
    \centering
    \includegraphics[width=\textwidth]{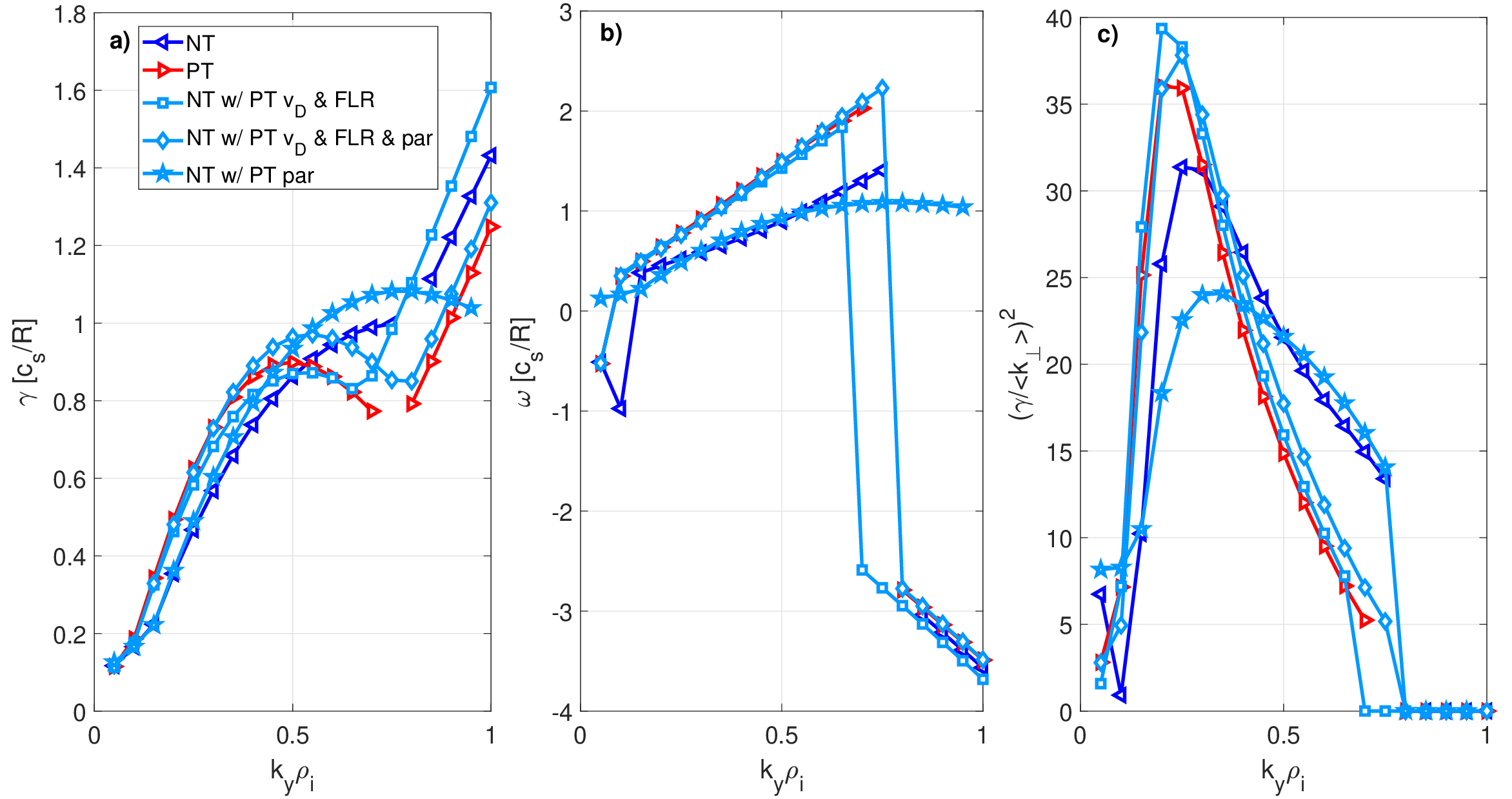}
    \caption{Same as figure~\ref{DIIID_gamma_omega_weight_vs_ky_swap} for AUG cases.}
    \label{AUG_gamma_omega_weight_vs_ky_swap}
\end{figure}

\begin{figure}[htbp]
    \centering
    \includegraphics[width=\textwidth]{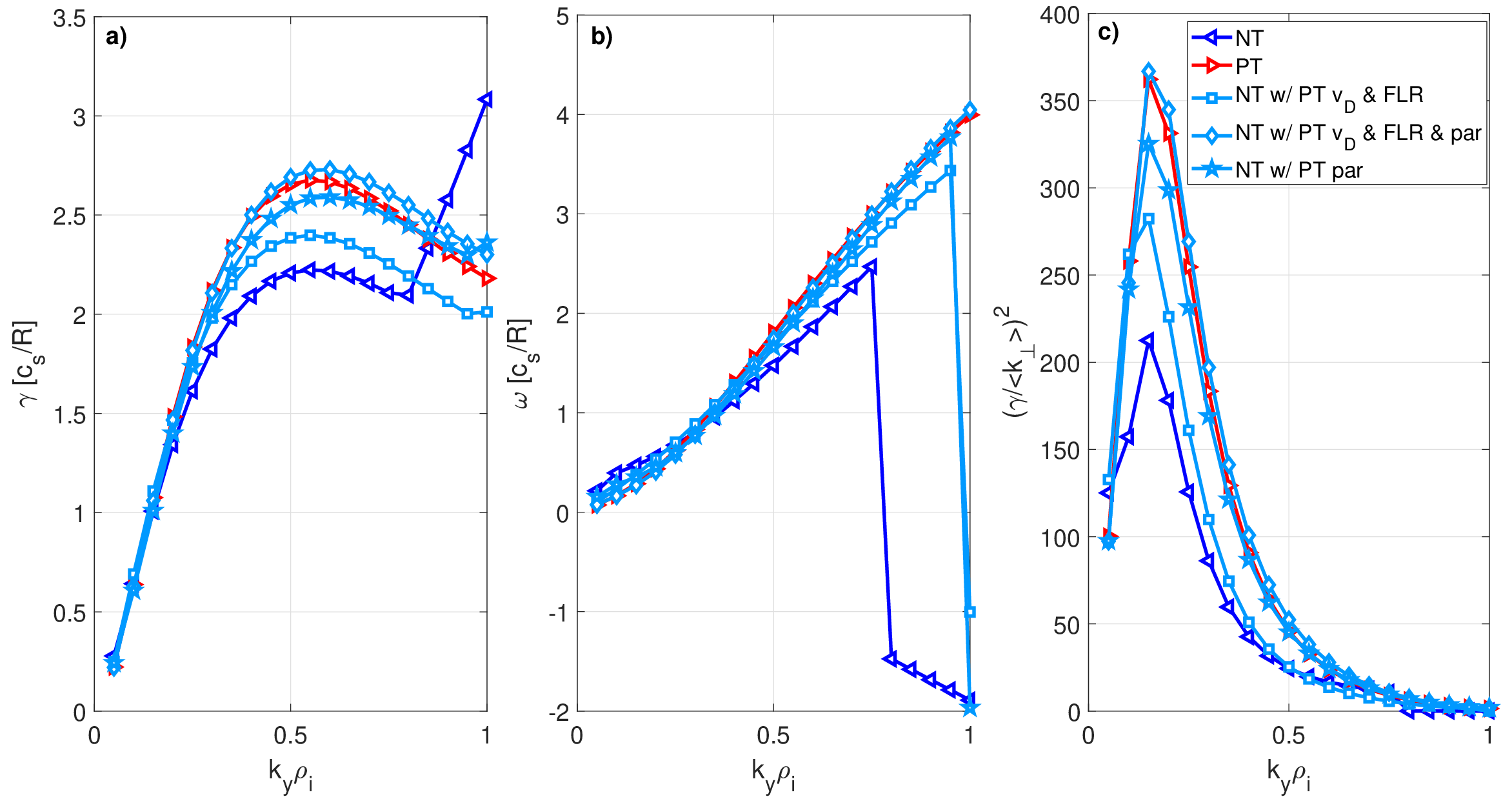}
    \caption{Same as figure~\ref{DIIID_gamma_omega_weight_vs_ky_swap} for JET cases.}
    \label{JET_gamma_omega_weight_vs_ky_swap}
\end{figure}

Linear simulations confirm that all three scenarios are dominated by ITG modes over the ion-scale range that contributes most to transport ($0.15 \lesssim k_y\rho_i \lesssim 0.5$), as demonstrated by the frequency spectra and QL weights shown in figures~\ref{DIIID_gamma_omega_weight_vs_ky_swap}, \ref{AUG_gamma_omega_weight_vs_ky_swap} and \ref{JET_gamma_omega_weight_vs_ky_swap}. All NT cases exhibit a TEM transition at high $k_y$ and, for AUG, microtearing modes at very low $k_y$. However, the QL analysis suggests that these modes contribute negligibly to transport. In all three cases the NT scenario is more linearly stable than PT in the transport-relevant part of the spectrum. The most unstable ITG growth rate is reduced by a factor of roughly 1.5 in DIII-D, with comparable reduction in JET and slightly decreased reduction in AUG. To isolate the contributions of individual geometric terms, we also systematically impose the PT geometric coefficients for magnetic drift velocities, FLR effects and parallel dynamics, following the procedure of \cite{Balestri_2024_2}. The results are shown in figures~\ref{DIIID_gamma_omega_weight_vs_ky_swap}, \ref{AUG_gamma_omega_weight_vs_ky_swap} and \ref{JET_gamma_omega_weight_vs_ky_swap}.
For DIII-D and AUG the picture is clear and consistent because imposing $v_{Dy}$ and FLR effects alone from PT into NT increases the growth rates more than swapping the parallel dynamics alone over the transport-relevant wavenumber range. For JET the linear results present a partial exception because swapping parallel dynamics alone has a somewhat larger impact on the growth rates than when swapping $v_{Dy}$ and $g^{yy}$. However, this will not be true in nonlinear simulations. Nonetheless, we remark that swapping all geometrical proxies together yields an almost perfect recovery of the linear growth rates of the self-consistent PT case in all three scenarios.

\subsection{Nonlinear simulations}

Nonlinear simulations were performed using the same parameter sets as the linear runs, with the box dimensions $L_x\sim180\rho_i$, $L_y\sim250\rho_i$, $256$ $k_x$ and $48$ $k_y$ modes, and $64$ parallel grid points.
\begin{figure}[htbp]
    \centering
    \begin{subfigure}
        {\includegraphics[width=\textwidth]{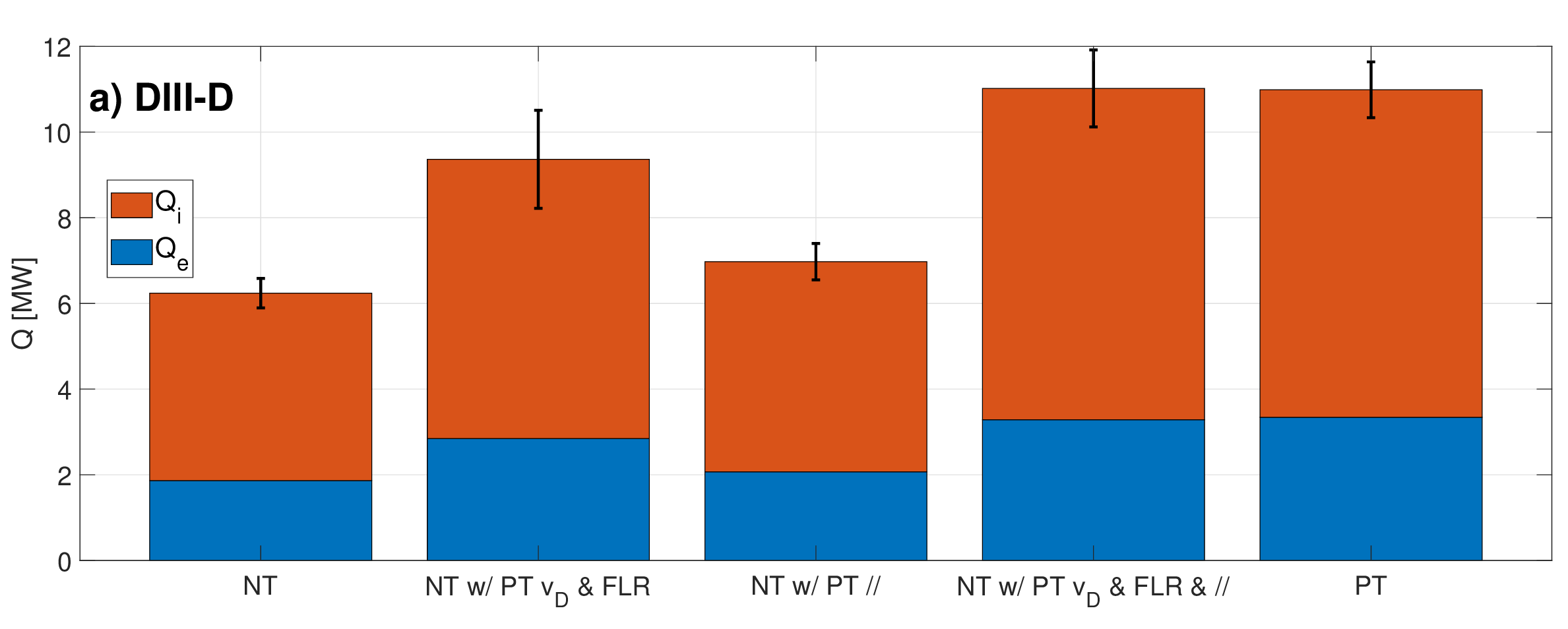}}
    \end{subfigure}
    \begin{subfigure}
        {\includegraphics[width=\textwidth]{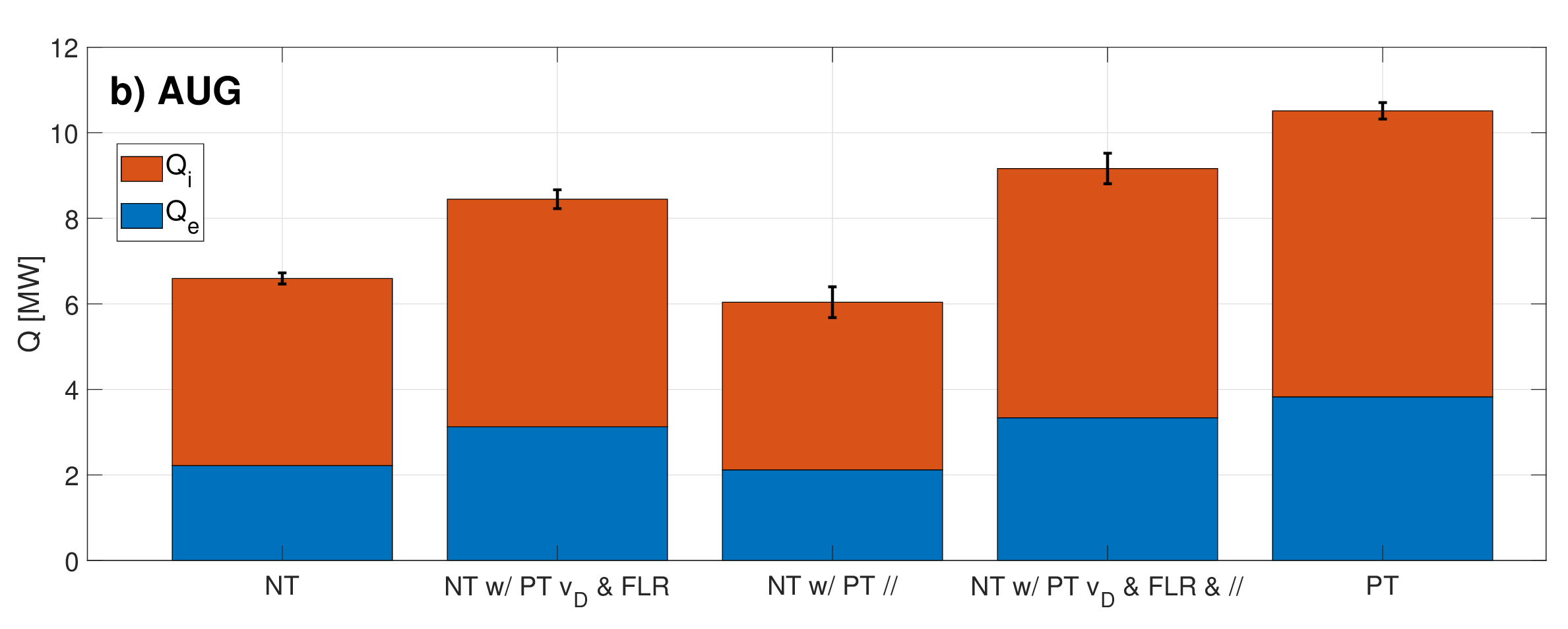}}
    \end{subfigure}
    \begin{subfigure}
        {\includegraphics[width=\textwidth]{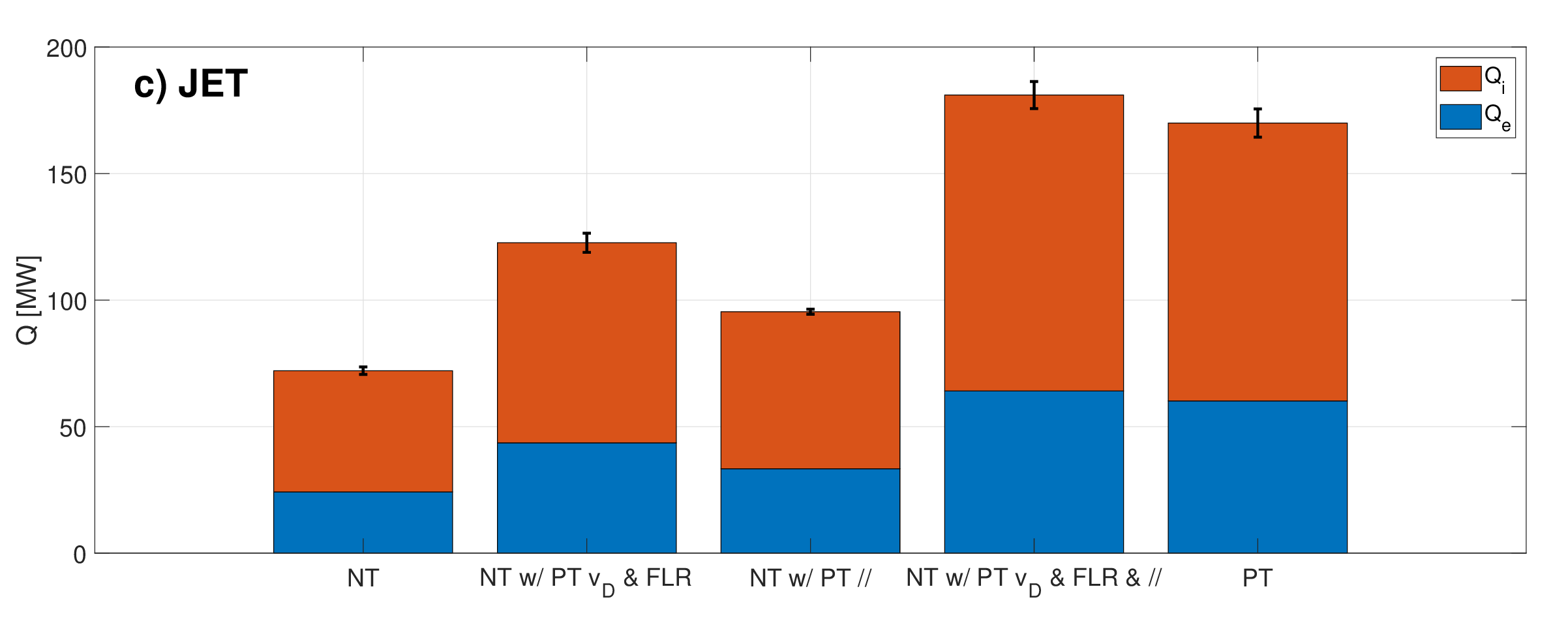}}
    \end{subfigure}
    \caption{Nonlinear heat fluxes for the coefficient-swap cases: electron electrostatic (blue) and ion electrostatic (orange) components for the (a) DIII-D, (b) AUG and (c) JET cases.}
    \label{DIIID_AUG_JET_Q_NL}
\end{figure}
The self-consistent NT and PT heat fluxes are compared in figures~\ref{DIIID_AUG_JET_Q_NL}. In all three machines NT features a substantial reduction of total heat flux with respect to PT: a factor of 1.8 for DIII-D, 1.6 for AUG and 2.1 for JET. The larger ion heat flux fraction in all scenarios confirms the ITG-dominated nature of the turbulence. For DIII-D and AUG, the NT heat fluxes are also in quantitative agreement with the total injected power reported at the simulated radial locations, supporting the reliability of the simulations. It is remarkable that the simulations predict a significant beneficial impact of NT also in AUG. For JET the heat flux is larger than the experimental estimate. This is due to our simulations neglecting impurities. We performed an additional simulation for the NT case including beryllium as an impurity and the heat flux significantly decreased, becoming comparable to the experimental heat flux. However, these simulations are significantly more expensive and the impact of impurities is outside the scope of this work. Nonetheless, we recognize the importance of impurities on transport and the eventual impact of geometry. This is left for future work.  Also for JET, which is the largest machine where a NT plasma has been produced, it is worth highlighting that a significant beneficial impact of NT on turbulence is found by GENE. The coefficient-swapping exercise with nonlinear simulations reinforces the linear picture. For DIII-D, imposing $v_D$ and FLR effects from PT increases the NT heat flux by a factor of 1.5, nearly recovering the PT level, while imposing only the parallel dynamics yields a much milder increase. For AUG, swapping $v_D$ and FLR effects increases the heat flux by a factor of 1.3, whereas imposing only the parallel dynamics produces no significant change. For JET, swapping $v_D$ and FLR effects increases the heat flux by a factor of 1.6 and swapping only parallel dynamics gives a smaller factor of 1.3, suggesting that nonlinear dynamics suppresses the role of parallel dynamics relative to what the linear analysis implies. In all three machines, swapping all coefficients together recovers the self-consistent PT heat flux.

\begin{figure}[htbp]
    \centering
    \includegraphics[width=\textwidth]{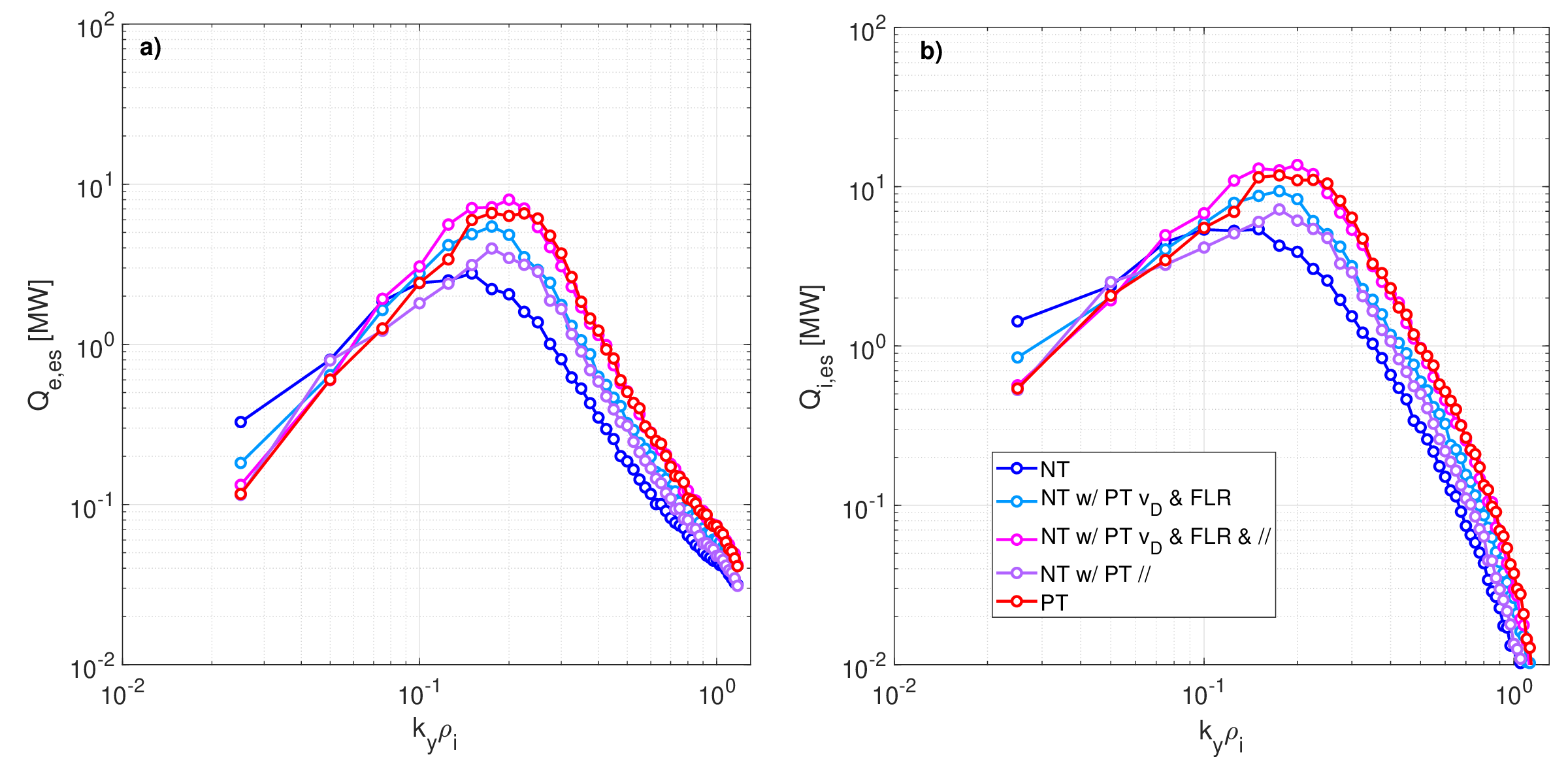}
    \caption{Electron (a) and ion (b) heat flux spectra for the different scenarios for JET.}
    \label{JET_Q_spectra}
\end{figure}

As done earlier, we can gain more insight by looking at the nonlinear spectra of the heat flux. For brevity, in figure \ref{JET_Q_spectra} we show only the spectra for the JET case, but the same picture holds for DIII-D and AUG. We observe that, consistent with what was predicted from the QL weights, the fluxes peak around $k_y\rho_i=0.2$. In agreement with the QL weights, we also see that the reduction of the heat flux in the NT case involves the whole ion scale, except for $k_y\rho_i<0.1$. All the artificial scenarios with swapped metrics show spectra that are very similar in shape to the PT ones. We see that the peak of the spectra moves to slightly larger $k_y$. As expected, the best agreement with the PT scenario is the NT case where we flip all the metrics together. This scenario reproduces quite well the full spectrum structure, which means that we are not matching the heat fluxes by chance, but we have almost identical systems in terms of turbulence. The other two cases are placed in between the PT and NT self-consistent scenarios, with the case with swapped $v_D$ and FLR effects being the closest to the spectrum of the PT scenario. 

\subsection{Summary}

Analysis of three tokamaks with different plasma shapes and experimental conditions indicates that the confinement improvement of NT over PT in ITG-dominated turbulence is primarily due to faster and more strongly varying magnetic drift velocities and by stronger FLR effects. Parallel dynamics plays a secondary role. This conclusion is in agreement with the results of the idealized pure-ITG cases discussed in \cite{Balestri_2024_2}, demonstrating that the metric-based framework provides reliable predictions of ITG transport differences across a broad range of scenarios.

In general, there are no matched PT counterparts in AUG and JET that can be used to directly experimentally assess the improvement of confinement coming from the NT geometry. Therefore, it is particularly noteworthy that simulations predict a significant beneficial impact of NT also in AUG and JET.

\section{Conclusions}\label{6}

In this work we extended and tested the physical picture based on geometric proxies developed in \cite{Balestri_2024_2} against experimental data and multi-machine gyrokinetic simulations, extending its applicability from simplified turbulent regimes to realistic tokamak conditions.

In the first part of the paper, we investigated the role of elongation in changing the beneficial effect of NT on turbulent transport with metric-based predictions, gyrokinetic simulations and experiments on TCV. The geometric coefficients predicted that the stabilizing effect of NT should diminish and eventually reverse as elongation is reduced below $\kappa\lesssim 0.75$, based on changes in the poloidally averaged magnetic drift velocity and FLR effects, as well as the variation of the magnetic drift profile along the poloidal angle. This prediction was confirmed by linear and nonlinear gyrokinetic simulations in both ITG- and TEM-dominated scenarios, which showed that NT features larger heat fluxes than PT for $\kappa<0.75$ and smaller heat fluxes for $\kappa>0.75$, in both turbulent regimes. The trend was validated experimentally through a dedicated elongation scan on TCV, where NT plasmas at high elongation ($\kappa^{\rm LCFS}=1.5$) achieved values of $\beta_N$ and $H_{98y2}$ approximately 1.5 times larger than their PT counterparts. Instead, at low elongation ($\kappa^{\rm LCFS}=0.95$) no appreciable difference in confinement between NT and PT was observed. This excellent agreement between geometrical proxy-based predictions, simulations and experiment constitutes strong evidence for the physical picture and demonstrates that our set of geometric proxies captures the essential physics behind the effect of plasma shape on turbulent transport.

In the second part of the paper, we extended the validation to ITG-dominated scenarios from DIII-D, AUG and JET, including plasmas with up-down asymmetric shapes that cannot be described using conventional triangularity. In all three cases, NT was found to exhibit significantly lower heat fluxes than the corresponding PT scenarios, with reductions of approximately 1.8, 1.6 and 2.1 times for DIII-D, AUG and JET respectively. By imposing the geometric coefficients of the PT cases into the NT simulations, we were able to disentangle the individual contributions of magnetic drifts, FLR effects and parallel dynamics to the NT stabilization. Consistently across all three machines, magnetic drifts and FLR effects emerged as the dominant stabilizing mechanisms, together accounting for most of the difference in transport between NT and PT. Parallel dynamics were found to play a secondary role in all the considered scenarios, with the partial exception of the JET case where its contribution was somewhat more pronounced in linear simulations, though less so nonlinearly. These findings confirm that the physical picture is robust and machine-independent, and that it remains valid even in realistic ITG-dominated conditions where the geometry is complex and the turbulence is not a pure single instability regime.

The results of this work demonstrate that simple geometric proxies based on the poloidal averages of the magnetic drift velocity, its poloidal variation and FLR effects represent reliable predictors of the impact of plasma shape on turbulent transport in ITG-dominated regimes, and also in TEM-dominated ones. Their predictive capability has been validated across a wide range of conditions, including up-down asymmetric plasma shapes, providing a general framework for the interpretation and optimization of NT scenarios. This work suggests that these proxies could serve as practical tools for the design and assessment of NT plasmas in future devices, enabling efficient screening of equilibria without the need for costly nonlinear gyrokinetic simulations.

\section*{Acknowledgments}
We acknowledge Stephan Brunner, Fabien Jeanquartier, Vincent Masson, Oleg Krutkin, Ralf Mackenbach and Haomin Sun for the constant help. This work has been carried out within the framework of the EUROfusion Consortium, via the Euratom Research and Training Programme (Grant Agreement No 101052200 — EUROfusion) and funded by the Swiss State Secretariat for Education, Research and Innovation (SERI). Views and opinions expressed are however those of the author(s) only and do not necessarily reflect those of the European Union, the European Commission, or SERI. Neither the European Union nor the European Commission nor SERI can be held responsible for them. This work was supported in part by the Swiss National Science Foundation. The authors acknowledge the CINECA award under the ISCRA initiative and the EUROfusion WP-AC, for the availability of high-performance computing resources and support. This research is supported by the U.S. Department of Energy under grants DE-SC0014264.

\bibliographystyle{unsrt}
\bibliography{refs}

\end{document}